\documentclass[
reprint,
superscriptaddress,
amsmath,amssymb,
aps,pre
]{revtex4-2}

\usepackage{amsfonts}

\usepackage{graphicx}% Include figure files
\usepackage{dcolumn}% Align table columns on decimal point
\usepackage{bm}% bold math
\usepackage{hyperref}% add hypertext capabilities
\usepackage{physics}
\usepackage{xcolor}
\usepackage{mathrsfs}
\usepackage{mathtools}
\usepackage{comment}
\begin{document}

\title{The long-range origin of the black hole entropy}

\author{Andrea Solfanelli}
\email{solfanelli@pks.mpg.de}
\affiliation{Max Planck Institute for the Physics of Complex Systems, Nöthnitzer Str. 38, 01187 Dresden, Germany}
\affiliation{SISSA, via Bonomea 265, 34136 Trieste, Italy}
%\affiliation{INFN, Sezione di Trieste, via Valerio 2, 34127 Trieste, Italy}
%\affiliation{Center for Life Center for Life Nano-Neuro Science @ La Sapienza, Italian Institute of Technology, 00161 Roma, Italy}

\author{Francesco Petocchi}
\affiliation{Department of Quantum Matter Physics, University of Geneva, 24 Quai Ernest-Ansermet, 1211 Geneva 4, Switzerland}

\author{Nicol\`o Defenu}
%\email{ndefenu@phys.ethz.ch}
\affiliation{Institut f\"ur Theoretische Physik, ETH Z\"urich, Wolfgang-Pauli-Str. 27 Z\"urich, Switzerland}

\date{\today} 
\begin{abstract}
%Reconciling quantum mechanics with gravitational physics remains one of the foremost challenges in modern theoretical physics, especially in the context of black hole thermodynamics and quantum chaos. The holographic principle offers a promising framework, suggesting that quantum gravity in a higher-dimensional bulk can be mapped to a non-gravitational system on its boundary. One such boundary system, 
The Sachdev-Ye-Kitaev (SYK) model, has emerged as a powerful tool for exploring the quantum nature of black holes, particularly their residual entropy at zero temperature. In this work, we investigate the role of long-range interactions in a chain of SYK dots with power-law decaying couplings and demonstrate how these interactions can lead to finite residual entropy in the strong non-local case. Indeed, as a function of the interaction range, the black-hole phase, found in the isolated SYK, melts into a long-range Fermi-liquid phase in the weak long-range regime and finally reaches the non Fermi-liquid phase in the purely local case. Our results pave the way to the understanding of the role of the interaction range in black hole thermodynamics and quantum information. \end{abstract}

\maketitle

One of the central challenges in modern physics is reconciling quantum mechanics with gravitational physics, classically governed by Einstein's general theory of relativity. When these two theories intersect, such as when describing the event horizons of black holes, paradoxes arise \cite{hawking1976breakdown}. One of the most promising approaches for addressing these contradictions is the holographic principle, which posits that quantum gravity in a $(d + 1)$-dimensional spacetime ``bulk" can be equivalently described by a non-gravitational many-body system defined on its $d$-dimensional boundary \cite{susskind1995theworld}. %This concept has been thoroughly realized in numerous examples of holographic dualities, offering a way to explore gravitational physics through the lens of quantum many-body systems, which are better understood \cite{franz2018mimicking}. 
The study of holographic quantum matter has revealed a deep and intriguing connection between the quantum properties of black holes and more conventional, yet strongly interacting, quantum many-body systems \cite{franz2018mimicking,hartnoll2018holographic}. This connection becomes evident when investigating their thermodynamic properties, statistical mechanics \cite{strominger1996microscopic}, and phenomena such as quantum chaos \cite{shenker2014blackholes,swingle2016measuring}. Classically, black holes are defined by the presence of an event horizon, a surface beyond which no information escapes. However, this defining feature of black holes necessitates revision from a quantum mechanical perspective. The discovery of Hawking radiation \cite{hawking1975particle} transformed our grasp of black hole thermodynamics. Hawking radiation is thermal, akin to that of a black body, indicating that black holes possess a temperature and exhibit behavior analogous to thermodynamic systems \cite{bekenstein1973blackholes,bardeen1973thefour}.

In this context the Sachdev-Ye-Kitaev (SYK) model\,\cite{sachdev1993gapless,kitaev2015simple,maldacena2016remarks} has recently gathered considerable attention both from the high energy and the condensed matter community. The SYK model is a quantum mechanical system with a large number of fermions ($N\gg1$) interacting via random all-to-all $q$-body couplings. This model exhibits a range of fascinating behaviors, including exotic Non-Fermi Liquid (NFL) states without quasiparticle excitations \cite{parcollet1999nonfermi,chowdhury2022syk,chowdhury2018translationally,banerjee2017solvable,haldar2018higher,jose2022nonfermi}, a connection to holographic dualities \cite{sachdev2015bekensteinhawking,cotler2017blackholes,susskind2021entanglement}, and quantum chaos \cite{shenker2014blackholes,maldacena2016abound,fu2016numerical}. Notably, through holographic duality, the SYK model has revealed similarities to black holes, providing a framework to probe the quantum nature of these enigmatic objects. A particularly important feature in this context is the black hole entropy, given by the Bekenstein-Hawking formula. In the case of charged black holes, particularly in near-horizon regions like $\mathrm{AdS}_2$ spacetime, a nonzero entropy persists even as the temperature approaches zero \cite{chamblin1999charged,faulkner2011emergent,sen2005blackhole,sen2008entropy,wald1993blackhole,jacobson1994onblackhole}. This behavior mirrors the entropy in the SYK model and reinforces the black hole analogy \cite{parcollet1998overscreened,parcollet2001quantum,sachdev2015bekensteinhawking}. The goal of this work is to explore how the residual entropy at zero temperature, a key aspect of this black hole SYK connection, originates from the long-range (LR) nature of the model interactions.

\begin{figure*}
    \centering    \includegraphics[width=\linewidth]{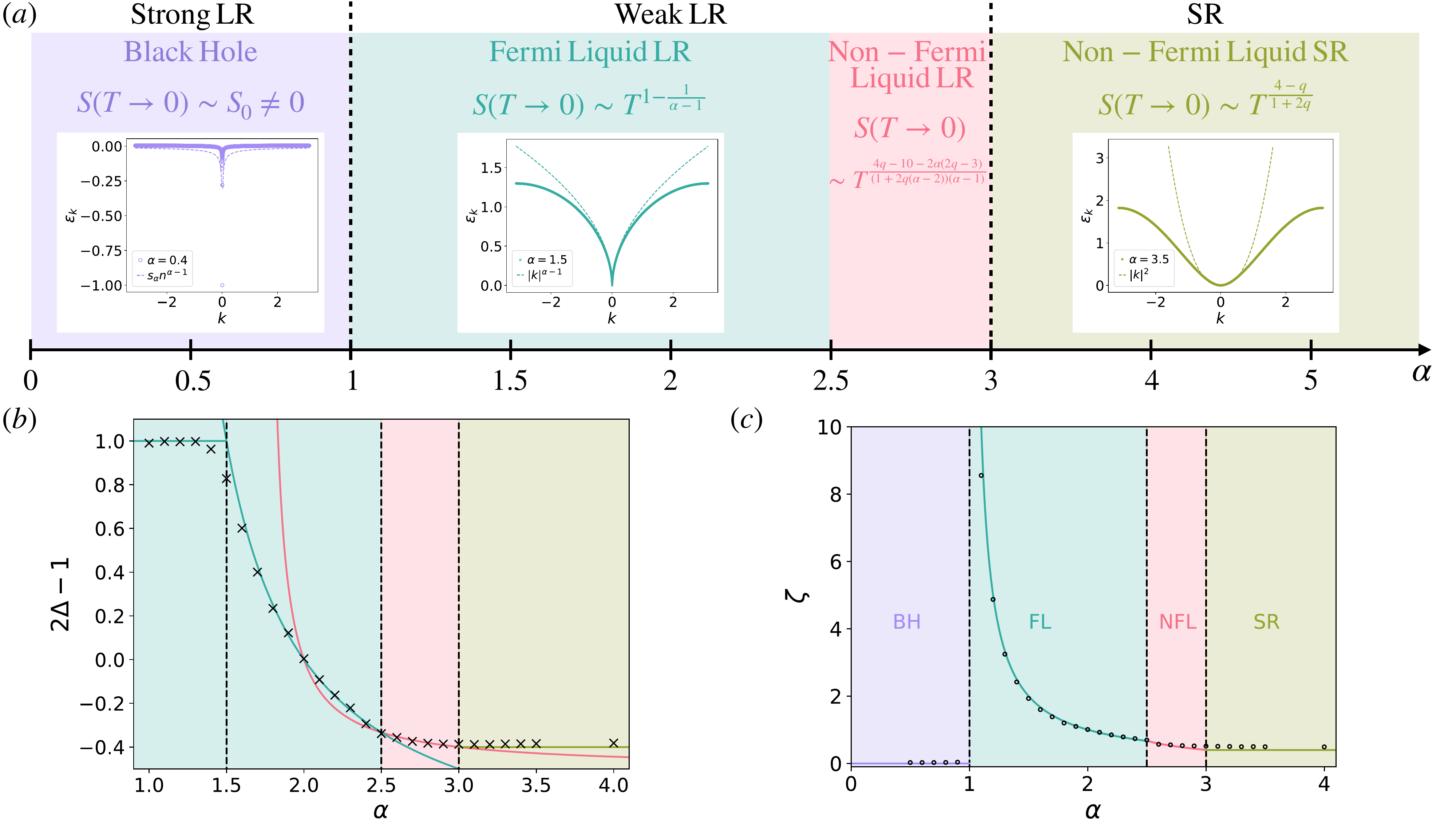}
    \caption{(a) Schematic phase diagram illustrating the different regimes as a function of the interaction range parameter $\alpha$ for a $d = 1$ chain of SYK dots with $q = 2$ body interactions. The black arrow indicates increasing values of $\alpha$, which delineate the Strong LR regime ($0<\alpha<d=1$), the Weak LR regime ($1<\alpha<3$) and the SR regime ($\alpha>3$). The insets depict the single-particle energy spectrum $\varepsilon_k = \varepsilon_\alpha(k)$ as a function of the Fourier modes $k$. The behavior of $\varepsilon_k$ near the zero mode ($k\approx 0$) and at the spectrum accumulation point $n\to\infty$ is captured through expansions represented by dashed lines, highlighting the different characteristics in the weak and strong LR regimes. The phases are color-coded, each representing a distinct zero-temperature entropy scaling behavior. (b) Numerical zero-temperature Green's function scaling exponent  $G(\omega)\approx C\omega^{2\Delta-1}$ (black crosses), plotted as a function of $\alpha$ and compared with the analytical predictions (solid lines). (c) The numerically obtained exponent $\zeta$, characterizing the low-temperature entropy scaling, $S(T\to 0)\approx S_0T^\zeta$  (black dots), is plotted as a function of $\alpha$ and compared with analytical predictions (solid lines). }
   \label{fig: SYK_Figure_bis}
\end{figure*}
In order to elucidate the mechanism underlying the presence of finite entropy at zero temperature, we introduce a hopping coupling between different SYK dots that decays as a power law with their distance $r$, i.e., $t_r\propto 1/r^\alpha$. In the following we refer to this model as the $\alpha$-SYK model. The system's behavior then strongly depends on the value of the power-law exponent $\alpha>0$, showing a reach phenomenology, as schematically summarized in Fig.~\ref{fig: SYK_Figure_bis}. When $\alpha>3$ the decay is rapid, effectively reducing the system to a short-range regime where the residual entropy vanishes at zero temperature, as it was demonstrated in the seminal work\,\cite{song2017strongly} for random local hopping amplitudes. Indeed, in the $\alpha>3$ limit, the SYK interactions only influence the scaling of the low-temperature entropy $S(T\to 0)\sim T^{\frac{4-q}{1+2q}}$ without generating any finite $T=0$ contribution. This phase goes beyond the description of standard Landau-Fermi liquid theory and is equivalent to the NFL phase found in the nearest-neighbor hopping ($\alpha\to\infty$) case in Refs. \cite{haldar2018higher,chowdhury2018translationally,jose2022nonfermi}.
For $q+1/2<\alpha<3$ the power law decaying couplings begin to influence the system low-energy behavior, resulting in a novel NFL phase where entropy displays an $\alpha$-dependent scaling exponent with the temperature going to zero $S(T\to 0)\sim T^{\frac{4q-10-2\alpha(2q-3)}{(1+2q(\alpha-2))(\alpha-1)}}$. In the regime $1<\alpha<q+1/2$, the system transitions to a Fermi liquid phase, where interactions become irrelevant at low energies, and the entropy scaling is tied to the single-particle dispersion relation $\varepsilon_\alpha(k)\approx |k|^{\alpha-1}$, leading to $S(T\to 0)\sim T^{1-\frac{1}{\alpha-1}}$. As
$\alpha\to 1$, the entropy exponent diverges, signaling a sharp change in behavior when $\alpha$ crosses the value of the system spatial dimension $d = 1$.
Finally, when $\alpha<1$, the system enters the strong LR regime, characteristic of non-additive systems dominated by LR couplings. Here, the residual entropy of the $\alpha$-SYK model remains finite, replicating the black hole-like behavior seen in an isolated SYK grain. This regime highlights the profound impact of LR interactions on the system low-energy properties and establishes the crucial connection between LR interactions and the appearance of finite residual entropy, a hallmark of black hole physics. This can be explained by general considerations on the peculiar properties of the spectrum of LR Hamiltonians\,\cite{defenu2021metastability}.
\section{The LR spectrum}
We begin by reviewing key spectral properties of LR Hamiltonians, which form the foundation for the distinct behaviors observed at different values of $\alpha$. A physical system is classified as LR when its coupling matrix $t_{ij}$ decays as a power law of the distance $r=|i-j|$ between its microscopic components $t_{ij}\propto r^{-\alpha}/\mathcal{N}_{\alpha}$ \cite{defenu2023longrange,defenu2024outofequilibrium}. The normalization factor $\mathcal{N}_{\alpha} = \sum_{r=1}^{L}r^{-\alpha}$ is introduced to ensure that the system internal energy remains extensive\,\cite{kac1963van}. In a $d$-dimensional system, depending on the value of the decay exponent $\alpha$, three different regimes can be identified:  
\begin{itemize}
    \item \textbf{SR regime} ($\alpha>\alpha^*$): The system's behavior mimics that of nearest-neighbor couplings.
    \item \textbf{The weak LR regime} ($d<\alpha<\alpha^*$): In this regime, thermodynamics is well-defined, but LR interactions significantly affect phase transitions and the universal scaling near classical and quantum critical points. The threshold value $\alpha^{*}$ depends on the specific model and phenomenon considered\,\cite{defenu2023longrange}.
    \item \textbf{The strong LR regime} ($0<\alpha<d$): The system energy is non-additive, and standard thermodynamics is not strictly valid \cite{campa2014physics}.
\end{itemize}

As a minimal model of a LR quantum system, we consider a generic system of particles hopping on a one-dimensional ($d=1$) lattice with LR, translationally invariant hopping amplitudes and possibly interacting among each other. The system is described by the Hamiltonian
\begin{align}
    \hat{H} = -\sum_{i=1}^L\sum_{r=1}^{L/2-1} t_r\left(\hat{c}^\dagger_i \hat{c}_{i+r}+h.c.\right)+\mu\sum_{i=1}^L\hat{c}^\dagger_i\hat{c}_i + \hat{H}_\mathrm{int}.\label{eq: first LR H}
\end{align}
where $\hat{c}^\dagger_i(\hat{c}_{i})$ are the creation (annihilation) operators for quantum particles at site $i$ along the chain, and $L$ is the total number of lattice sites. The nature of the particles (whether bosons or fermions) and the specific form of the interaction Hamiltonian $\hat{H}_{\mathrm{int}}$ are not crucial at this stage.

We begin by diagonalizing the non-interacting part of the Hamiltonian. Under the assumption of periodic boundary conditions, the spectrum of the non-interacting Hamiltonian is given by $\varepsilon_\alpha(k) = \mu - f_{\alpha}(k)$, where
\begin{align}
    f_{\alpha}(k) =  \frac{1}{\mathcal{N}_\alpha}\sum_{r = 1}^{L/2-1}\frac{\cos(kr)}{r^{\alpha}},\label{eq: falpha}
\end{align}
is the Fourier transform of the LR hopping amplitudes $t_r$ and we have the usual restriction on the momentum, imposed by the periodic boundary conditions, $k\equiv k_n = 2\pi n/L$ with $n\in \mathbb{Z}$ and $n = \lfloor -L/2\rfloor,\dots\lfloor L/2\rfloor$ (the lattice spacing has been set to 1). %The behavior of $f_\alpha(k)$ for different values of $\alpha$, plays a crucial role for understanding the physics of LR quantum systems, making it important to review its properties. 
%First of all we notice that the Kac normalization $\mathcal{N}_\alpha$ scales differently with the system size $L\gg 1$ depending on $\alpha$. In the strong LR regime $\alpha<1$, the normalization factor is diverging with the system size as $\mathcal{N}_\alpha\approx L^{1-\alpha}/c_\alpha$, with $c_\alpha = (1-\alpha)2^{1-\alpha}$. Instead, in the weak LR case \alpha>1
%\ln L &\mathrm{if}\,\,\alpha = 1\\
%        \zeta(\alpha) &\mathrm{if}\,\,\alpha>1
%    \end{cases},\label{eq: kac scaling}
%\end{align}
%where $c_\alpha = (1-\alpha)2^{1-\alpha}$ and $\zeta(x)$ is the Riemann zeta function.

As long as we are in the weak LR regime $\alpha>1$, the normalization factor $\mathcal{N}_\alpha$ converges in the thermodynamic limit $\mathcal{N}_{\alpha}\underset{L\to\infty}{\longrightarrow}\zeta(\alpha)$
% $L\to\infty$ limit, i.e., $\lim_{L\to\infty}\mathcal{N}_\alpha = \zeta(\alpha)$
, where $\zeta(x)$ is the Reimann zeta function. As a result, the calculation proceeds similarly to the nearest-neighbor case, allowing us to replace the discrete momentum values $k_n$ with a continuous variable $k\in[-\pi,\pi)$. %In this regime, the spectrum of the Hamiltonian becomes continuous, and $f_\alpha(k)$ takes the form \cite{defenu2019universal}
%\begin{align}
%    f_\alpha(k)\approx \frac{1}{\zeta(\alpha)}\left[\mathrm{Li}_\alpha(e^{ik})+\mathrm{Li}_\alpha(e^{-ik})\right]\label{eq: falpha weak LR}
%\end{align}
%where $\mathrm{Li}_x(z) = \sum_{n = 1}^{\infty}z^n/n^x$ is the polylogarithm \cite{abramowitz1964handbook}.
While the specific lattice type may affect the exact form of $f_\alpha(k)$, the essential physics of LR interacting systems is dominated by its asymptotic behavior at low energy. Of particular interest are the low $k$ modes of the single particle spectrum, which determine the dispersion relation of the LR system. By expanding Eq.\,\eqref{eq: falpha} at leading order around $k = 0$, the $\alpha$-dependent dispersion relation, at the gapless point, becomes $\varepsilon_\alpha(k)\approx |k|^{\alpha-1}$, as long as $\alpha<3$, while for $\alpha>3$, we recover the standard nearest-neighbor tight-binding dispersion relation $\varepsilon_\alpha(k)\approx |k|^{2}$. Once the dispersion relation is known, we can compute the low-energy density of states as
\begin{align}
    g(\varepsilon) = \int_{-\pi}^\pi \frac{dk}{2\pi}\delta(\varepsilon-\varepsilon_\alpha (k)) \approx\begin{cases}|\varepsilon|^{\frac{1}{\alpha-1}-1} &1<\alpha<3\\
    |\varepsilon|^{-1/2}&\alpha> 3
    \end{cases},\label{eq: density of states}
\end{align}
%This result can be compared to the standard power-law scaling of the density of states in a local system, allowing us to define the spectral dimension, $d_s$,  through $g(\varepsilon)\approx \varepsilon^{d_s/2-1}$ \cite{bighin2024universal,hattori1987gaussian}. In the case of LR interactions, the spectral dimension is then given by $d_s = 2/(\alpha-1)$. 
For the case under scrutiny, the threshold at which short-range spectral properties are recovered corresponds to  $\alpha^* = 3$\,\cite{defenu2024outofequilibrium}.%., corresponds to the value of $\alpha$ at which the spectral dimension matches the actual spatial dimension of the system.

In the strong LR regime, where $\alpha < 1$, the situation changes dramatically as the normalization factor $\mathcal{N}_\alpha$ diverges with $L$. This divergence ensures that the system energy remains extensive, but it also requires careful consideration when taking the thermodynamic limit of the Fourier transform $f_\alpha(k)$. %To handle this, we express Eq.\,\eqref{eq: falpha} for large $L$ as 
%\begin{align}
 %   \lim_{L\to\infty}\frac{1}{\mathcal{N}_\alpha}\sum_{r = 1}^{L/2-1}\frac{\cos(kr)}{r^\alpha}\approx\frac{c_\alpha}{L}\sum_{r = 1}^{L/2}\frac{\cos(2\pi n \frac{r}{L})}{(r/L)^\alpha}.
%\end{align}
%In this expression, 
In particular, as detailed in the Supplemental Material, the $1/L$ scaling of the momenta implies that the summation in Eq.\,\eqref{eq: falpha} depends only on the variable $r /L$. Therefore, in the thermodynamic limit ($L\to\infty$), we can approximate the sum as an integral over the continuous variable $s = r/L$, yielding
\begin{align}
    f_\alpha(n) = \lim_{L\to\infty}f_\alpha(k) = \int_0^{1/2} ds\frac{\cos(2\pi n s)}{s^\alpha}.\label{eq: falpha spectrum}
\end{align}
Despite its simplicity, this result has profound physical implications: it shows that, in the strong LR regime, the spectrum of a quantum system with LR harmonic couplings remains discrete even as $L\to\infty$. %Specifically, for $\alpha<1$, the energy gap between consecutive eigenvalues $\varepsilon_{n+1}-\varepsilon_n$ (labeled by consecutive momenta $k_n$, $k_{n+1}$) in Eq.\,\eqref{eq: falpha}, does not vanish in the thermodynamic limit, contrary to the behavior observed for $\alpha>1$. 
Indeed, the energy eigenvalues $\varepsilon_n = \mu - f_\alpha(n)$ depend only on the integer index  $n\in \mathbb{Z}$ rather than on a continuous momentum variable $k$. %Notably, for $\alpha = 0$, we find that  $f_{\alpha = 0}(n) = \delta_{n,0}$, leading to a fully degenerate discrete spectrum with $\varepsilon_n = \mu$ for $n\neq 0$ and $\varepsilon_n = \mu-1$ for $n=0$. 
Additionally, the energy eigenvalues are not densely distributed. Each eigenvalue is isolated, with the only accumulation point at the maximum energy $\max_n\varepsilon_n = \mu$. This follows from the Riemann–Lebesgue lemma \cite{last1996quantum}, which implies $\lim_{n\to\infty}f_\alpha(n) = 0$. More precisely close to the spectrum accumulation point $n\to\infty$ we have that $f_\alpha(n)\approx s_\alpha n^{\alpha-1}$.
\section{The $\alpha$-SYK model}
\begin{figure}
    \centering    \includegraphics[width=\linewidth]{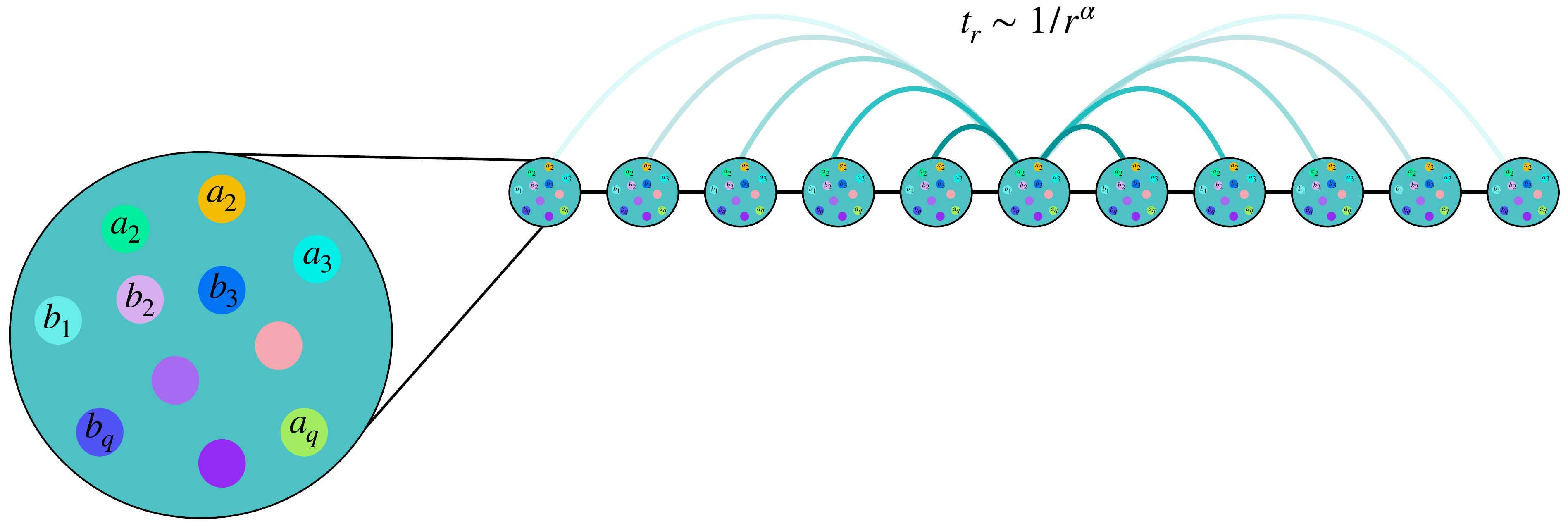}
    \caption{Schematic representation of the $\alpha$-SYK model: a lattice of SYK dots connected through power law decaying hopping amplitudes.}
   \label{fig: SYK_LR_disegno}
\end{figure}
With this understanding of the single-particle spectrum in LR systems, we are now ready to introduce interactions into the model. Specifically, we focus on the case of fermionic SYK interactions. As illustrated in Fig.\,\ref{fig: SYK_LR_disegno}, we consider a one-dimensional ($d = 1$) lattice of length $L$, where each site hosts an SYK dot. The system consists of $N$ flavors of complex fermions, each hopping between the lattice sites with power-law-decaying hopping amplitudes $t_r = t\,r^{-\alpha}/\mathcal{N}_\alpha$. Each SYK dot at site $x$ has intradot complex random all-to-all 
$q$-body interactions, with the couplings $U_{a_1,\dots a_q;b_1,\dots b_q}^x$ drawn from a Gaussian distribution with zero mean $\langle U_{a_1,\dots a_q;b_1,\dots b_q}^x\rangle = 0$ and  with variance $\langle |U_{a_1,\dots a_q;b_1,\dots b_q}^x|^2\rangle = U^2/qN^{2q-1}(q!)^2$. The Hamiltonian of the $\alpha$-SYK model then takes the general form of Eq.\,\eqref{eq: first LR H}, with the interaction term given by
\begin{align}
    \hat{H}_\mathrm{int} =\sum_{x = 1}^L\sum_{\substack{a_1,...a_q\\b_1,...b_q}}^NU_{a_1,...a_q; b_1,... b_q}^x\hat{c}_{a_1,x}^\dagger...\hat{c}_{a_q,x}^\dagger \hat{c}_{b_1,x}...\hat{c}_{b_q,x},\label{eq:LR SYK Hamiltonian}
\end{align}
where, the indices $a_i,b_j = 1,\dots N$ label the $N$ fermionic flavors at each site, and $x=1,\dots,L$ runs over the lattice sites. 

In the limit of large number of fermionic flavors $N\to\infty$, the saddle-point approximation becomes exact and the model can be treated analytically. To derive the saddle-point equations, we first perform a disorder average over the random SYK couplings using the replica method. This results in an effective action that can be simplified within a replica-diagonal ansatz, expressed in terms of a large-$N$ collective field $G_{x}(\tau_1,\tau_2)$ and its conjugate $\Sigma_x(\tau_1,\tau_2)$, where $\tau_{1,2}$ are imaginary times. At the saddle point, $G$ is the on-site fermion Green’s function and $\Sigma$ is the self-energy. By extremizing the effective action and assuming color symmetry and translational invariance, we obtain the following self-consistent equations for $G$ and $\Sigma$
\begin{align}
    \Sigma(\tau) &= (-1)^{q+1}U^2G^q(\tau)G^{q-1}(-\tau),\label{eq: saddle point equation}\\
    G(i\omega_n) &= \frac{1}{L}\sum_{k}\frac{1}{i\omega_n-\varepsilon_k-\Sigma(i\omega_n)},\label{eq: green function}
\end{align}
where $\omega_n = (2n+1)\pi/\beta$ are the fermionic Matsubara frequencies 
%such and we have assumed time 
implying time translation invariance $G(\tau_1,\tau_2) = G(\tau_1-\tau_2)$. At zero temperature ($T = 0$), we can obtain the low-energy solution to these equations analytically. This requires distinguishing between two cases: the weak LR regime ($\alpha>1$) and the strong LR regime ($\alpha<1$).
\subsection{Weak LR}
When $\alpha > d = 1$, the single-particle spectrum becomes continuous in the thermodynamic limit, allowing us to safely introduce the density of states as defined in Eq.\,\eqref{eq: density of states}. Then, performing the analytic continuation $i\omega_n \to \omega + i0^{+}$, we can express Eq.\,\eqref{eq: green function} as an integral over the continuous energy spectrum (see the Methods section for further details). In this regime, the saddle point equations admit two possible low-energy solutions \cite{haldar2018higher}: an interaction-dominated solution for $\omega \ll \Sigma(\omega)$ as $\omega \to 0$, and a hopping-dominated solution for $\Sigma(\omega) \ll \omega$ as $\omega \to 0$. The latter corresponds to a Fermi liquid fixed point where interactions become irrelevant as $\omega \to 0$, while the former describes a non-trivial fixed point, leading to a NFL behavior. 

Let us first consider the second case. Analogously to the case of an isolated SYK grain, the solutions for $G(\omega)$ and $\Sigma(\omega)$ can be sought using the power-law ansatz $G(\omega) = C e^{i\theta} \omega^{2\Delta - 1}$ and $\Sigma(\omega) = C_\Sigma e^{i\theta_{\Sigma}} \omega^{2\Delta_{\Sigma} - 1}$, and inserting them into the low-energy expansion of the Saddle Point Equations\,\eqref{eq: saddle point equation} and\,\eqref{eq: green function}. From this, we find that $\Delta_{\Sigma} = (2q - 1)\Delta$, and we can extract the $\alpha$ dependence of the fermionic scaling dimension 
\begin{align}
    \Delta = \begin{cases}
        \frac{3}{(2 + 4 q)} &\mathrm{\alpha>3}\\
        \frac{2\alpha-3}{2[1+2q(\alpha-2))]} &\alpha_c(q)<\alpha< 3
    \end{cases}\label{eq: Delta NFL}
\end{align}
In order to estimate the threshold value $\alpha_c(q)$ the assumption $\omega \ll \Sigma(\omega)$ as $\omega \to 0$ has to be self consistently verified. This yields the condition for a NFL phase, $2\Delta_{\Sigma} - 1 \leq 1$, which translates into the following condition for the power-law decay exponent
\begin{align}
    \alpha\leq\alpha_c(q) = \frac{1}{2} + q
\end{align}
We notice that since $q>1$ then $\alpha_c(q)>3/2$ and in particular $\alpha_c(q=2) = 5/2$.

Below the $q$-dependent critical value $\alpha_c(q)$, the NFL scaling solution no longer exists. This brings us to the saddle-point solution for the perturbative fixed point, which corresponds to the regime where $\omega \gg \Sigma(\omega)$. In this case, it can be shown that the fermionic scaling dimension $\Delta$ behaves as
\begin{align}
    \Delta = \begin{cases}
        \frac{1}{2(\alpha-1)} &\frac{3}{2}<\alpha\leq\alpha_c(q)\\
        1 &1<\alpha\leq\frac{3}{2}\label{eq: Delta FL}
    \end{cases}
\end{align}
Where the changing in the scaling at the $\alpha = 3/2$ point is due to a change in the Green's function low energy expansion (see the Methods section for additional details).

Figure \ref{fig: SYK_Figure_bis}(b) illustrates the Green's  function scaling exponent $2\Delta - 1$, for the $\alpha$-SYK model at zero temperature, as a function of $\alpha$, demonstrating excellent agreement between the analytical results in Eqs.\,\eqref{eq: Delta NFL} and\,\eqref{eq: Delta FL} and the full numerical solutions of Eqs.\,\eqref{eq: saddle point equation} and\,\eqref{eq: green function}.
\subsection{Strong LR}\label{sec: Strong LR}
In the strong LR regime, where $0 < \alpha < 1$, the single-particle spectrum becomes discrete even in the thermodynamic limit and it is better described by the discreted index $n$: $\lim_{L \to \infty} \varepsilon_k = \varepsilon_n$. As a result, we cannot take the continuum limit, and the density of states as defined in Eq.\,\eqref{eq: density of states} is no longer well-defined. Consequently, the sum over Fourier modes in Eq.\,\eqref{eq: green function} transforms into a sum over the discrete labels $n$. 

The leading contribution to this sum arises from values of $n$ near the accumulation point, $n \to \infty$, of the discrete spectrum. Expanding the spectrum near this accumulation point as $\varepsilon_n \approx s_\alpha n^{\alpha - 1}$ in Eq.\,\eqref{eq: green function} and retaining only the leading order contribution as $n \to \infty$, we obtain
\begin{align}
    G(\omega) \approx \frac{1}{\omega-\Sigma(\omega)} +\mathcal{O}(L^{\alpha-1}).\label{eq: green function strong LR}
\end{align}
This implies that the Green's function of the $\alpha$-SYK model in the strong LR regime follows the same equation as for a single SYK grain without any hopping contribution, apart from finite size corrections which are subleading in the $L\to\infty$ limit.
\section{The residual entropy}
\begin{figure}
    \centering    \includegraphics[width=0.9\linewidth]{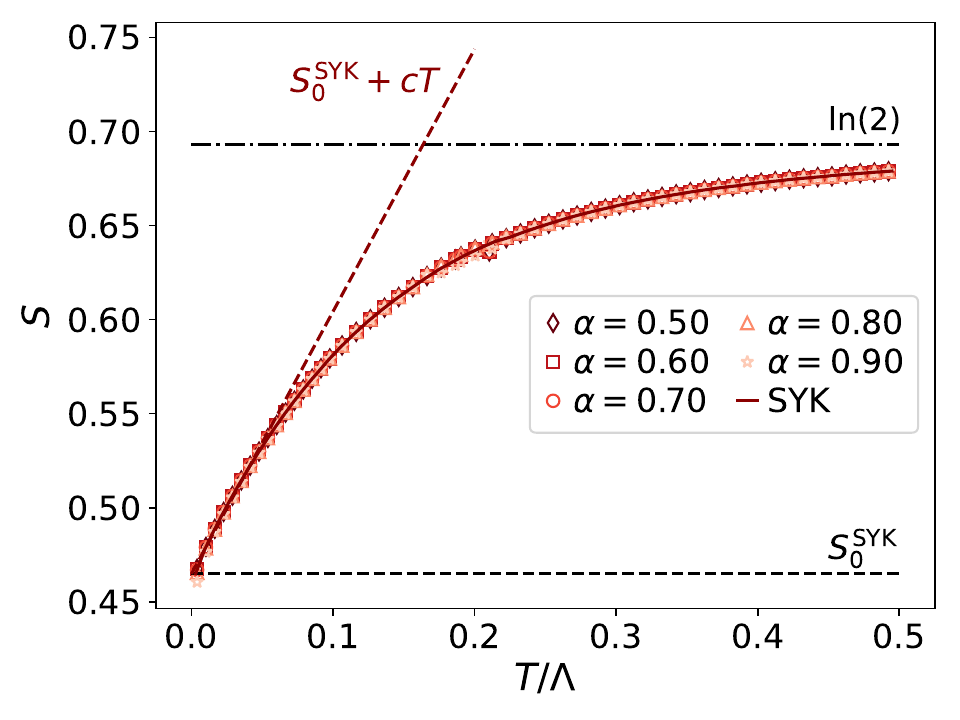}
    \caption{Entropy density for the $\alpha$-SYK model as a function of temperature $T$ (in units of the bandwidth $\Lambda$) for various values of $\alpha$ in the strong LR regime, $\alpha < 1$ (scatter plots), compared with entropy density of the zero dimensional SYK model (dark red solid line). The black dashed line represents the single SYK grain prediction for the residual entropy at zero temperature, while the dark red dashed line indicates the first linear correction at finite temperatures. The dot-dashed line shows the infinite temperature entropy, $\ln(2)$, which the system approaches as $T \to \Lambda$.}
   \label{fig: ResidualEntropy_StrongLR}
\end{figure}
To compute the system entropy, we first need to extend the results of the previous section to finite temperature. Analytically, the isolated SYK grain can be solved at finite temperature by exploiting the invariance of the saddle point equations under any imaginary time reparametrization, $\tau \to \sigma(\tau)$. This allows us to obtain the finite temperature result, $T = 1/\beta$, by applying the mapping $\tau = (\beta/\pi) \tan(\sigma \pi/\beta)$ to the zero-temperature solution. However, this conformal symmetry is broken in the $\alpha$-SYK model when the hopping term is introduced. Nevertheless, as shown in Ref.\,\cite{haldar2018higher}, assuming approximate time reparametrization invariance still leads to excellent quantitative agreement with numerical results as long as $\alpha>\alpha_c(q)$. Crucially, this approximation allows us to extract the entropy scaling exponent, $\zeta$, in the $T \to 0$ limit of the $\alpha$-SYK model. In the weak LR regime, this approach connects the exponent $\zeta$ to the fermionic scaling dimension $\Delta$ in Eq.\,\eqref{eq: Delta NFL}. Specifically, in the NFL regime where $\alpha > \alpha_c(q)$, we find
\begin{align}
    \zeta = \frac{(2(2q-1)\Delta-1)}{\alpha-1}.
\end{align}
While, in the Fermi liquid phase, where $1 < \alpha \leq \alpha_c(q)$, the entropy exponent is determined solely by the singularity in the dispersion relation, leading to
\begin{align}
    \zeta = \frac{1}{\alpha-1}.
\end{align}

In contrast, as described in Section \ref{sec: Strong LR}, the situation changes entirely in the strong LR regime, where $0 < \alpha < 1$. In this case, the saddle point solutions of the $\alpha$-SYK model become equivalent to those of a single SYK grain, with only subleading corrections vanishing as $L \to \infty$. Remarkably, this results in the finite residual entropy at zero temperature
\begin{align}
    S_0^{\alpha-\mathrm{SYK}} = S_0^{\mathrm{SYK}} + \mathcal{O}(L^{\alpha-1}).
\end{align}
where $S_0^{\mathrm{SYK}}$ is the residual entropy of the single grain SYK model\,\cite{sachdev2015bekensteinhawking,kitaev2015simple,maldacena2016remarks}. Therefore, in the strong LR regime $\alpha < d$, the system enters the black hole phase, where the zero temperature entropy remains finite.

These predictions are numerically benchmarked by directly solving Eqs.\,\eqref{eq: saddle point equation} and\,\eqref{eq: green function} at finite temperature. The system entropy density, $S$, is obtained using the relation $S = -\partial F/\partial T$, where $F$ represents the free energy density evaluated at the saddle point. The numerical results in Fig.\,\ref{fig: SYK_Figure_bis}(c) are obtained by fitting the entropy density data as a function of temperature, $T$, with the power-law dependence $S(T \to 0)=S_0 T^\zeta$. This fit shows excellent agreement with our theoretical predictions clearly showing a singular behavior of the exponent $\zeta$ as the parameter $\alpha$ approaches the critical value $\alpha = d = 1$. Specifically, we observe $\lim_{\alpha\to 1^+}\zeta = +\infty$, while $\zeta = 0$ for all $\alpha<1$ signaling the emergence of the black-hole phase as the notion of locality is lost.

Furthermore, Fig.\,\ref{fig: ResidualEntropy_StrongLR} shows the numerical results for the entropy density blue of the $\alpha$-SYK model as a function of temperature (measured in units of the bandwidth $\Lambda$) for different values of $\alpha<1$ confirming a strong agreement with the behavior of a single SYK grain. As discussed in Section \ref{sec: Strong LR}, this is consistent with the argument that in the strong LR regime, the system exhibits the same low energy properties as the single SYK grain, up to subleading finite-size corrections. In particular, at sufficiently low temperatures, the entropy converges to the residual value $S_0^\mathrm{SYK}$ and the leading-order finite-temperature corrections follow the expected linear dependence on $T$.

\section{Discussion}
Our results pioneers the examination of how non-locality affects the scaling of thermodynamic quantities in strongly correlated and AdS/CFT models and particularly the emergence of finite residual entropy at zero temperature. The power-law decay of hopping couplings, in the $\alpha$-SYK model, introduces a rich phenomenology, with distinct regimes characterized by different entropy scaling laws. For rapid decay ($\alpha>3$) the system behaves like a SR model, where the residual entropy vanishes as temperature approaches zero, though with non-trivial NFL scaling corrections. As the interaction range increases, novel NFL behavior emerges for $q+1/2<\alpha<3$ driven by the long-range interactions. In this regime, the entropy scaling becomes $\alpha$-dependent, with the system showing clear deviations from standard Fermi liquid theory. In contrast, when 
$1<\alpha<q+1/2$, the system transitions to a Fermi liquid phase, and the entropy scaling is determined by the single-particle spectrum. Finally in the strong LR regime ($\alpha<d = 1$), the system enters the black hole phase where the entropy at zero temperature is finite. This demonstrates a profound connection between the notion of locality and the thermodynamic properties of black hole analogs. These findings underscore the unique role of long-range interactions in shaping low-energy quantum states and provide new avenues for exploring quantum gravity through condensed matter systems.
%\newpage
\section{Methods}
\subsection{Low energy expansion of the zero temperature Green function}
In the weak LR regime, where $1 < \alpha < 3$, the system allows for the introduction of the density of states\,\eqref{eq: density of states}. Performing the analytic continuation $i\omega_n\to\omega+i0^{+}$, we can rewrite Eq.\,\eqref{eq: green function} as
\begin{align}
 G(\omega)\approx G(z) = \int_{-\Lambda}^{\Lambda}d\varepsilon \frac{g(\varepsilon)}{z-\varepsilon},
\end{align}
where we have introduced the quantity $z = \omega-\Sigma(\omega)$ and $\Lambda$ is the bandwidth. By applying the low-energy expansion of the density of states, Eq.\,\eqref{eq: density of states}, and retaining only the leading-order term as  $z\to 0$  we derive the following expansion for the Green function
\begin{align}
    G(z)\propto \begin{cases} 
    z^{-\frac{1}{2}} &\alpha\geq 3\\
    z^{-1+\frac{1}{\alpha-1}} & \frac{3}{2}<\alpha<3\\
    z &1<\alpha\leq \frac{3}{2}
    \end{cases},\label{eq: G(w) equations weak LR}
\end{align}
where the change in the scaling behavior of the Green function at $\alpha = 3/2$ underpins the change in the fermionic scaling dimension discussed in Eq.\,\eqref{eq: Delta FL} (see Supplemental Material for further details).

On the other hand, in the strong LR regime, where $0<\alpha<1$, the expression for the green function of Eq.\,\eqref{eq: green function} becomes
\begin{align}
    G(\omega) \approx \frac{1}{L}\sum_{n}\frac{1}{z-\varepsilon_n},\label{eq: green function strong LR}
\end{align}
where the spectrum is discrete and labeled by the integer $n$. Expanding the spectrum near its accumulation point and focusing on the leading-order contribution as $n\gg 1$, we obtain
\begin{align}
    G(\omega) \approx \frac{1}{z}\left[1-\frac{s_\alpha}{zL}\sum_{n}\frac{1}{n^{1-\alpha}}+\mathcal{O}(L^{-1})\right],
\end{align}
This result reduces to Eq.,\eqref{eq: green function strong LR} in the thermodynamic limit, $L\to\infty$, where the sum over $n$ gives the sumleading contribution $\frac{1}{L}\sum_n n^{\alpha-1} = \mathcal{O}(L^{\alpha-1})$.

\subsection{Finite temperature numerics}
Finite temperature calculation were performed by solving Eq.\eqref{eq: saddle point equation} and Eq.\eqref{eq: green function} within the Matsubara finite temperature formalism. The algorithm starts with the evaluation of the non-interacting Green's function $G_0$:
\begin{align}
    G_0\left( i\omega_n \right) = \int \frac{g\left(\varepsilon \right) d\varepsilon}{ i\omega_n - \varepsilon}
\end{align}
where we numerically compute the density of states $g\left(\varepsilon \right)$ for each value of the $\alpha$ parameter. To capture the long-range behavior of the kinetic energy $\varepsilon_{\alpha}\left(k \right)$ we retain real-space hopping up to the 10$^5$-th neighbor while sufficient energy resolution was attained with 10$^8$ k-points. For $\alpha>1$ finite-size effects in $g\left(\varepsilon \right)$ become increasingly evident as we move deeper into the LR regime. This is due to the growing scarcity of states in the lower bottom of the single particle spectrum and further increasing the number of k-points only marginally mitigates this numerical artifact. In order to compensate for this effect, we exploit the analytical low-energy formulations of Eq.\eqref{eq: density of states} to replace the portion of the density of states with finite-size effects (ad-hoc constants were fitted to ensure $\mathcal{C}^2$ continuity with the $g\left(\varepsilon \right)$ obtained numerically). Starting from the initial $G_0$ we self-consistently update Eq.\eqref{eq: saddle point equation} and Eq.\eqref{eq: green function}. Convergence is achieved when the Neumann distance between self-energies is less than 10$^{-6}$. Calculations have been performed considering $U=t=\mu=1$ and a Matsubara frequency cutoff of $1500t$. We verified that increasing this cutoff does not appreciably affect the high-temperature results. Free energy was computed via direct implementation of Eq. 46 of the SM.

%
%\begin{align}
%f_\alpha(k) &= 1+\sin\left(\frac{\alpha\pi}{2}\right)\frac{\Gamma(1-\alpha)}{\zeta(\alpha)}|k|^{\alpha-1}+\mathcal{O}(k^2) &\mathrm{if}\,\,1<\alpha<3,\\
%f_\alpha(k) &=  1+\frac{2\ln(k)-3}{4\zeta(3)}k^2 + \mathcal{O}(k^3) &\mathrm{if}\,\,\alpha = 3,\\
%f_\alpha(k) &=  1-\frac{\zeta(\alpha-2)}{2\zeta(\alpha)}k^2 + \mathcal{O}(k^{\alpha-1}) &\mathrm{if}\,\,\alpha > 3.
%\end{align}
%

\section{Acknowledgements} 
This research was supported in part by grant NSF PHY-230935 to the Kavli Institute for Theoretical Physics (KITP). N.D. acknowledges funding by the Swiss National Science Foundation (SNSF) under project funding ID: 200021 207537 and by the Deutsche Forschungsgemeinschaft (DFG, German Research Foundation) under Germany’s Excellence Strategy EXC2181/1-390900948 (the Heidelberg STRUCTURES Excellence Cluster).
%%%%%%%%%%%%%%%%%%%

\begin{widetext}
    
\begin{center}
     \textbf{\Large Supplementary information: The long-range origin of the black hole entropy}
\end{center}

\section{Expansions of the long-range spectrum}
In this section we provide the details on the long-range single particle spectrum $\varepsilon_\alpha(k) = \mu - f_{\alpha}(k)$ properties. In particular we focus on the peculiar properties of the Fourier transform of the LR hopping coupling
\begin{align}
    f_{\alpha}(k) =  \frac{1}{\mathcal{N}_\alpha}\sum_{r = 1}^{L/2-1}\frac{\cos(kr)}{r^{\alpha}},\label{eq: falpha}
\end{align}
in the different regimes identified by different values of $\alpha$. As stated in the main text, such regimes can be identified by noticing that the Kac normalization $N_\alpha$ scales differently with the system size $N\gg 1$ depending on $\alpha$:
\begin{align}
    N_\alpha\approx \begin{cases}
        N^{1-\alpha}/c_\alpha &\mathrm{if}\,\,\alpha<1\\
        \ln N &\mathrm{if}\,\,\alpha = 1\\
        \zeta(\alpha) &\mathrm{if}\,\,\alpha>1
    \end{cases},\label{eq: kac scaling}
\end{align}
where $c_\alpha = (1-\alpha)2^{1-\alpha}$ and $\zeta(x)$ is the Riemann zeta function and we are considering a long-range system  in one spacial dimension $d = 1$.

In the weak long-range regime $\alpha>1$, the Kac scaling is finite in the $N\to\infty$ limit. The thermodynamic limit of Eq.\,\eqref{eq: falpha} can be taken safely, and the single particle spectrum becomes continuous with $f_\alpha(k)$ reading \cite{defenu2019universal}
\begin{align}
    f_\alpha(k)\approx \frac{1}{\zeta(\alpha)}\left[\mathrm{Li}_\alpha(e^{ik})+\mathrm{Li}_\alpha(e^{-ik})\right]\label{eq: falpha weak LR}
\end{align}
where $\mathrm{Li}_x(z) = \sum_{n = 1}^{\infty}z^n/n^x$ is the polylogarithm \cite{abramowitz1964handbook}.

Then the low energy contributions close to the zero mode, these are obtained by taking the Taylor expansion of Eq.\,\ref{eq: falpha weak LR} around $k = 0$ leading to \cite{defenu2019universal}
\begin{align}
f_\alpha(k) &= 1+\sin\left(\frac{\alpha\pi}{2}\right)\frac{\Gamma(1-\alpha)}{\zeta(\alpha)}|k|^{\alpha-1}+\mathcal{O}(k^2) &\mathrm{if}\,\,1<\alpha<3,\\
f_\alpha(k) &=  1+\frac{2\ln(k)-3}{4\zeta(3)}k^2 + \mathcal{O}(k^3) &\mathrm{if}\,\,\alpha = 3,\\
f_\alpha(k) &=  1-\frac{\zeta(\alpha-2)}{2\zeta(\alpha)}k^2 + \mathcal{O}(k^{\alpha-1}) &\mathrm{if}\,\,\alpha > 3.
\end{align}
Then, at the gapless point $\mu = 1$, we find the $\alpha$ dependent dispersion relation $\varepsilon_\alpha(k)\approx |k|^{\alpha-1}$, as long as $\alpha<3$, while we retrieve the standard dispersion relation for a nearest neighbor tight binding model 
$\varepsilon_\alpha(k)\approx |k|^{2}$ when $\alpha>3$.

The situation changes dramatically in the strong long-range regime $\alpha < 1$. Indeed, as shown in Eq.\,\eqref{eq: kac scaling}, the Kac normalization factor $N_\alpha$ diverges at large $N$ ensuring energy extensivity. Accordingly, the thermodynamic limit of Eq.\,\eqref{eq: falpha} must be carefully considered. To this aim, it is convenient to write Eq.\,\eqref{eq: falpha} explicitly for large $N$ as 
\begin{align}
    \lim_{N\to\infty}\frac{1}{N_\alpha}\sum_{r = 1}^{N/2-1}\frac{\cos(kr)}{r^\alpha}\approx\frac{c_\alpha}{N}\sum_{r = 1}^{N/2}\frac{\cos(2\pi n \frac{r}{N})}{(r/N)^\alpha}.
\end{align}
Due to the $1/N$ scaling of the discrete momenta on the lattice, the summation depends only on the variable $r /N$. Therefore, for $N\to\infty$, we can take the continuum limit by transforming the sum over $r$ into an integral with respect to $s = r/N$, leading to
\begin{align}
    f_\alpha(n) = \lim_{N\to\infty}f_\alpha(k) = \int_0^{1/2} ds\frac{\cos(2\pi n s)}{s^\alpha}.\label{eq: falpha spectrum}
\end{align}
For $\alpha = 0$, we find that  $f_\alpha(n)\to\delta_{n,0}$, leading to a fully degenerate discrete spectrum: $\varepsilon_n = \mu$ for $n\neq 0$ and $\varepsilon_n = \mu-1$ for $n=0$. Moreover, the spectrum has only an accumulation point occurring at the maximum energy $\max_n\varepsilon_n = \mu$. This follows from the Riemann–Lebesgue lemma \cite{last1996quantum}, which implies 
\begin{align}
    \lim_{n\to\infty}f_\alpha(n) = 0.
\end{align}
More precisely, we can expand $f_\alpha(n)$ at leading order for $n$ close to the accumulation point $n\to\infty$, leading to
\begin{align}
    f_\alpha(n) =  n^{\alpha-1} (2\pi)^{\alpha-1}\Gamma(1-\alpha)\sin(\alpha\pi/2) + \mathcal{O}(n^{-2}) = s_\alpha n^{\alpha-1}+ \mathcal{O}(n^{-2}).
\end{align}
\section{Derivation of the saddle-point equations}
The SYK model with LR hopping is exactly solvable at the level of saddle point in the limit $N\to\infty$, i.e., when there are
a large number of SYK flavors. To derive the saddle-point equations we start from the Euclidean-time action in terms of Grassmann variables $(\bar{c},c)$
\begin{align}
    S = \int_0^\beta d\tau \sum_{x = 1}^L\sum_{a = 1}^N\left[\bar{c}_{x,a}\partial_{\tau}c_{x,a}-H(\bar{c},c)\right],
\end{align}
where $\beta = 1/T$ is the inverse temperature. Then, the disorder averaged action can then be written down using the replica-trick as
\begin{align}
    \mathcal{S} = &\int_0^\beta d\tau_1\int_0^\beta d\tau_2\left[\sum_{\nu_1}\sum_{x=1}^L\sum_{r=1}^{L/2-1}\sum_{a = 1}^N\bar{c}_{x,a,\nu_1}(\tau_1)[\partial_{\tau_1}\delta_{x,x+r}+t_r]\delta(\tau_1-\tau_2)c_{x+r,a,\nu_1}(\tau_2)\right.\notag\\
    &\left.-(-1)^q\frac{J^2}{2qN^{2q-1}}\sum_{\nu_1,\nu_2}\sum_{x = 1}^L\left(\sum_{a,b = 1}^N\bar{c}_{x,a,\nu_1}(\tau_1)c_{x,a,\nu_2}(\tau_2)\bar{c}_{x,a,\nu_2}(\tau_2)c_{x,a,\nu_1}(\tau_1)\right
)^q\right],
\end{align}
where $\nu_1,\nu_2$ are the replica indices. We introduce the large-$N$ field $G_{x,\nu_1,\nu_2}(\tau_1,\tau_2) = \frac{1}{N}\sum_{a =1}^N\bar{c}_{a x\nu_2}(\tau_2)c_{ax\nu_1}(\tau_1)$ and the Lagrange multiplier $\Sigma_{x,\nu_1,\nu_2}(\tau_1,\tau_2)
$ to obtain
\begin{align}
    \mathcal{S} = &\int_0^\beta d\tau_1\int_0^\beta d\tau_2\left[\sum_{\nu_1}\sum_{x=1}^L\sum_{r=1}^{L/2-1}\sum_{a = 1}^N\bar{c}_{x,a,\nu_1}(\tau_1)[\partial_{\tau_1}\delta_{x,x+r}+t_r]\delta(\tau_1-\tau_2)c_{x+r,a,\nu_1}(\tau_2)\right.\notag\\
    &\left.-N\sum_{x = 1}^L\sum_{\nu_1,\nu_2}\left[(-1)^q\frac{U^2}{2q}G_{x,\nu_1,\nu_2}^q(\tau_1,\tau_2)G_{x,\nu_2,\nu_1}^q(\tau_2,\tau_1)-G_{x,\nu_1,\nu_2}(\tau_1,\tau_2)\Sigma_{x,\nu_2,\nu_1}(\tau_2,\tau_1)\right]\right].
\end{align}
The kinetic part of the action can be diagonalized by passing to Fourier space defining $c_{k a \nu_1} = \frac{1}{\sqrt{L}}\sum_{x = 1}^L c_{x a \nu_1}e^{-ikx}$. Next, imposing lattice translational invariace, such that $G_x = G,\Sigma_x = \Sigma$, and using the replica-diagonal ansatz, we obtain
\begin{align}
    \mathcal{S} &= N\int_0^\beta d\tau_1\int_0^\beta d\tau_2\left[\sum_{k,a}\bar{c}_{k,a}(\tau_1)\left[(\partial_{\tau_1}+\varepsilon_k)\delta(\tau_1-\tau_2)+\Sigma(\tau_1,\tau_2)\right]c_{k,a}(\tau_2)\right.\notag\\
    &\left.-(-1)^q\frac{U^2}{2q}G^q(\tau_1,\tau_2)G^q(\tau_2,\tau_1)-\Sigma(\tau_2,\tau_1)G(\tau_1,\tau_2)\right],\label{eq: action}
\end{align}
where $\varepsilon_{k} = 1-t_k$ is the dispersion relation. Tracing over the fermionic degrees of freedom we get
\begin{align}
    \frac{\mathcal{S}}{N} = \int_0^\beta d\tau_1\int_0^\beta d\tau_2\left[-\sum_{k}\mathrm{Tr}\ln[(\partial_{\tau_1}+\varepsilon_k)\delta(\tau_1-\tau_2)+\Sigma(\tau_1,\tau_2)]-(-1)^q\frac{U^2}{2q}G^q(\tau_1,\tau_2)G^q(\tau_2,\tau_1)-\Sigma(\tau_2,\tau_1)G(\tau_1,\tau_2)\right].
\end{align}
By extremizing the action, we obtain the saddle-point equations for $G$ and $\Sigma$ as
\begin{align}
    &\Sigma(\tau_1,\tau_2) = (-1)^{q+1}U^2G^q(\tau_1,\tau_2)G^{q-1}(\tau_2,\tau_1),\\
    &G(\tau_1,\tau_2) = \langle\bar{c}_x(\tau_2)c_x(\tau_1)\rangle = \frac{1}{L}\sum_{k}G_k(\tau_1,\tau_2),
\end{align}
where
\begin{align}
    G_k(\tau_1,\tau_2) = -\frac{1}{(\partial_{\tau_1}+\varepsilon_k)\delta(\tau_1-\tau_2)+\Sigma(\tau_1,\tau_2)}.
\end{align}
Finally, using time-translational invariance, $G(\tau_1,\tau_2) = G(\tau_1-\tau_2)$, we obtain the equations
\begin{align}
    \Sigma(\tau) &= (-1)^{q+1}U^2G^q(\tau)G^{q-1}(-\tau),\label{eq: saddle point equation}\\
    G(i\omega_n) &= \frac{1}{L}\sum_{k}\frac{1}{i\omega_n-\varepsilon_k-\Sigma(i\omega_n)},\label{eq: green function}
\end{align}
where $\omega_n = (2n+1)\pi/\beta$ are the fermionic Matsubara frequencies such that
\begin{align}
    f(\tau) = \frac{1}{\beta}\sum_nf(i\omega_n)e^{-i\omega_n\tau},\quad f(i\omega_n) = \int_0^\beta d\tau f(\tau)e^{i\omega_n\tau}.
\end{align}

In the weak long-range regime $1 < \alpha < 3$, we can introduce the density of states $g(\varepsilon)$. It is then convenient to employ this density of states to pass from the sum over Fourier modes in Eq.\,\eqref{eq: green function} to an integral over $\varepsilon$, resulting in
\begin{align}
G(i\omega_n)\approx\int_{-\Lambda}^{\Lambda}d\varepsilon\frac{g(\varepsilon)}{i\omega_n-\varepsilon-\Sigma(i\omega_n)},\label{eq: green function density of states}
\end{align}
where we have introduced the band width $\Lambda$ and the density of states $g(\varepsilon)$.
\section{Zero temperature solutions}
In this section we provide the details of the calculation of the low energy solutions of the saddle point equations at zero temperature in the different $\alpha$ regimes.
\subsection{Weak long-range and short-range regimes}
For $\alpha>1$, we can safely introduce the density of states, which at low energy can be approximated as 
\begin{align}
    g(\varepsilon) &\approx \begin{cases}
       \frac{1}{(\alpha-1)}\left[-\frac{\zeta(\alpha)}{(\sin\left(\alpha\pi/2\right)\Gamma(1-\alpha)}\right]^{\frac{1}{\alpha-1}}|\varepsilon|^{-1+\frac{1}{\alpha-1}} &|\varepsilon|\leq\Lambda\\
       0 &|\varepsilon|>\Lambda
    \end{cases} &\mathrm{for}\quad 1<\alpha<3,\\
    g(\varepsilon) &\approx \begin{cases}
       \sqrt{\frac{\zeta(\alpha)}{2\zeta(\alpha-2)}}|\varepsilon|^{-\frac{1}{2}} &|\varepsilon|\leq\Lambda\\
       0 &|\varepsilon|>\Lambda
    \end{cases} &\mathrm{for}\quad \alpha> 3.
\end{align}
Accordingly, performing the analytic continuation $i\omega_n\to\omega+i0^{+}$, we can rewrite Eq.\,\eqref{eq: green function density of states} as
\begin{align}
 G(\omega)\approx\int_{-\Lambda}^{\Lambda}d\varepsilon \frac{g_\alpha |\varepsilon|^{-\gamma_\alpha}}{\omega-\Sigma(\omega)-\varepsilon},
\end{align}
where 
\begin{align}
    \gamma_\alpha = \begin{cases}
        1-1/(\alpha-1)\\\\
        1/2
    \end{cases}g_\alpha = \begin{cases}
        \frac{1}{(\alpha-1)}\left[-\frac{\zeta(\alpha)}{(\sin\left(\alpha\pi/2\right)\Gamma(1-\alpha)}\right]^{\frac{1}{\alpha-1}} &\mathrm{for}\quad1<\alpha<3\\\\
        \sqrt{\frac{\zeta(\alpha)}{2\zeta(\alpha-2)}} &\mathrm{for}\quad\alpha\geq3
    \end{cases}\label{eq: gamma_alpha g_alpha}
\end{align} 
Then, introducing the quantity $z = \omega-\Sigma(\omega)$, and performing the change of variables $y = \varepsilon/z$  we obtain
\begin{align}
     G(\omega)\approx g_\alpha z^{-\gamma_\alpha}\int_{-\Lambda/z}^{\Lambda/z}dy \frac{|y|^{-\gamma_\alpha}}{1-y} = g_\alpha z^{-\gamma_\alpha}\left[\int_{0}^{\Lambda/z}dy \frac{y^{-\gamma_\alpha}}{1-y}+\int_{0}^{\Lambda/z}dy \frac{y^{-\gamma_\alpha}}{1+y}\right],\label{eq: green function integral}
\end{align}
In the low energy regime $|z|\to 0$, the integrals in the previous expression can be expanded as follows
\begin{align}
    \int_{0}^{\Lambda/z}dy \frac{y^{-\gamma_\alpha}}{1\pm y} = (\pm 1)^{\gamma_{\alpha}+1}\frac{\pi}{\sin(\pi\gamma_{\alpha})}\pm\frac{z^{\gamma_{\alpha}}}{\Lambda^{\gamma_{\alpha}}\gamma_{\alpha}}+\frac{z^{1+\gamma_{\alpha}}}{\Lambda^{1+\gamma_{\alpha}}(1+\gamma_{\alpha})}+\mathcal{O}(z^2).
\end{align}
Inserting, this expansion into Eq.\eqref{eq: green function integral} and keeping only the leading order contributions we obtain
\begin{align}
    G(\omega)\approx \begin{cases}
    \frac{\pi g_\alpha(1-e^{i\pi\gamma_\alpha})}{\sin(\pi\gamma_\alpha)}z^{-\gamma_{\alpha}} &\gamma_{\alpha}>-1,\alpha>3/2\\\\
    \left[\frac{\pi g_\alpha(1-e^{i\pi\gamma_\alpha})}{\sin(\pi\gamma_\alpha)}+\frac{2g_\alpha }{\Lambda^{1+\gamma_\alpha}(1+\gamma_\alpha)}\right]z &\gamma_{\alpha}=-1, \alpha=3/2\\\\
    \frac{2g_\alpha }{\Lambda^{1+\gamma_\alpha}(1+\gamma_\alpha)}z &\gamma_{\alpha}<-1, 1<\alpha<3/2
    \end{cases}.\label{eq: G(w) equations weak LR}
\end{align}
In this regime, the saddle point equations allow for two possible solutions \cite{haldar2018higher}: an interaction-dominated solution for $\omega\ll\Sigma(\omega)$ as $\omega\to 0$, and an hopping dominated solution for $\omega\ll\Sigma(\omega)$ as $\omega\to 0$. The latter corresponds to a Fermi Liquid fixed point where the interaction becomes
irrelevant for $\omega\to 0$. The former is instead a non-trivial fixed point leading to a non-Fermi-liquid (NFL). Let us start from the second situation in which we set $z\approx-\Sigma(w)$ in Eq.\eqref{eq: G(w) equations weak LR}.
\begin{itemize}
    \item \emph{ $\alpha>\alpha_c(q)$:} In this regime the saddle point equations to be solved are
\begin{align}
    G(\omega) = (-1)^{\gamma_\alpha}\frac{\pi g_\alpha(1-e^{i\pi\gamma_\alpha})}{\sin(\pi\gamma_\alpha)}\Sigma(\omega)^{-\gamma_{\alpha}},\quad \Sigma(\tau) = (-1)^{q+1}G^q(\tau)G^{q-1}(-\tau).\label{eq: saddle point 1alpha1.5}
\end{align}
As in the conventional SYK model, at $T = 0$, the solution is obtained by taking a
power-law ansatz for $G(\omega)$, leading to
\begin{align}
    G(\omega) = Ce^{i\theta}\omega^{2\Delta-1},\quad \Sigma(\omega) = U^2\frac{C^{2q-1}}{\pi^{2q-1}}\frac{(\Gamma(2\Delta)\sin\theta)^{2q-1}}{\Gamma(2(2q-1)\Delta)\sin((2q-1)\Delta\pi)}e^{-i\pi(2q-1)\Delta}\omega^{2(2q-1)\Delta-1}
\end{align}
Then, the self-consistency condition of Eq.\,\eqref{eq: saddle point 1alpha1.5} fixes the fermion scaling dimension $\Delta$ and the prefactor $C$ to be
\begin{align}
    &\Delta = \frac{1+\gamma_\alpha}{2(1-\gamma_\alpha+2\gamma_\alpha q)} =\begin{cases}
        \frac{3}{(2 + 4 q)} &\mathrm{\alpha>3}\\\\
        \frac{2\alpha-3}{2[1+2q(\alpha-2))]} &\alpha_c(q)<\alpha\leq 3\end{cases},\label{eq: Delta NFL}\\\notag\\
    &C = \left[\frac{g_\alpha\pi^{2\gamma_\alpha(q-1)+1}}{U^{2\gamma_\alpha}\cos(\gamma_\alpha\pi/2)}\left(\frac{\Gamma(2\Delta_\Sigma)\sin(\pi\Delta_\Sigma)}{(\Gamma(2\Delta)\sin(\pi\Delta))^{2q-1}}\right)^{\gamma_\alpha}\right]^\frac{2\Delta}{1+\gamma_\alpha},
\end{align}
where $\Delta_\Sigma = (2q-1)\Delta$. Moreover, as shown in Ref. \cite{haldar2018higher} as long as $\gamma_\alpha<1$, which is always the case in our setup since $\gamma_\alpha<1/2$, then the low energy saddle point equations and the constraint $\mathrm{Im}G(\omega)<0$ completely fix the phase $\theta = -\pi\Delta$, without allowing any particle hole asymmetry.
The expression for the scaling dimension $\Delta$ then allow us to identified the critical value $\alpha_c(q)$ below which the condition $\omega\ll\Sigma(\omega)$ as $\omega\to 0$ is no more satisfied. This is achieved by imposing the condition
\begin{align}
    \Delta_\Sigma = (2q-1)\Delta = \frac{(2q-1)(2\alpha_c(q)-3)}{2[1+2q(\alpha_c(q)-2))]} = 1,
\end{align}
which immediately leads to
$\alpha_c(q) = 1/2+q$. We notice that since $q>1$ then $\alpha_c(q)>3/2$ and in particular $\alpha_c(q=2) = 5/2$.
\item \emph{$3/2<\alpha<\alpha_c(q)$:}  In this case the low energy limit of the saddle point equations still reads
\begin{align}
    G(z) \approx \frac{\pi g_\alpha(1-e^{i\pi\gamma_\alpha})}{\sin(\pi\gamma_\alpha)}z^{-\gamma_{\alpha}},\quad \Sigma(\tau) = (-1)^{q+1}U^2G^q(\tau)G^{q-1}(-\tau).
\end{align}
But now we are in the Fermi liquid phase with $z\approx\omega$ accordingly we find that
\begin{align}
    G(\omega)\approx\omega^{-{2\Delta-1}}\approx\omega^{\frac{1}{\alpha-1}-1},
\end{align}
leading to
\begin{align}
    \Delta = \frac{1}{2(\alpha-1)}.
\end{align}
\item $1<\alpha<3/2$: In this regime the saddle point equations to be solved become
\begin{align}
    G(z)\approx\frac{2g_\alpha }{\Lambda^{1+\gamma_\alpha}(1+\gamma_\alpha)}z,\quad\Sigma(\tau) = (-1)^{q+1}U^2G^q(\tau)G^{q-1}(-\tau)
\end{align}
Moreover, also in this case we are in a Fermi Liquid regime so that at low energy $z\approx \omega$ and we find that
\begin{align}
    G(\omega)\approx \omega^{2\Delta-1}\approx\omega,
\end{align}
leading to $\Delta = 1$.
\end{itemize}
\subsection{Strong long-range regime}
In the strong long-range regime 
$0<\alpha<1$, the single particle is descrete also in the thermodynamic limit, and labelled by the integer number $l$, i.e., $\lim_{L\to\infty}\varepsilon_k = \varepsilon$. As a consequence we cannot perform a continuum limit and the density of states $g(\varepsilon)$ is no more well defined. After the analytic continuation The expression for the green function of Eq.\,\eqref{eq: green function} then becomes
\begin{align}
    G(\omega) \approx \frac{1}{L}\sum_{l}\frac{1}{\omega-\Sigma(\omega)-\varepsilon_l} = \frac{1}{zL}\sum_{l}\frac{1}{1-\varepsilon_l/z},\label{eq: green function strong LR}
\end{align}
where, as in the previuos, section we have introduced the quantity $z = \omega-\Sigma(\omega)$ in the last identity. 

As detailed in Section \ref{sec: The model} spectrum can be well approximated by Eq.\,\eqref{eq: spectrum strong LR}. Accordingly, substituting the expansion in Eq.\,\eqref{eq: spectrum strong LR} in the expression for the green function in Eq.\,\eqref{eq: green function strong LR} we obtain
\begin{align}
    G(\omega) \approx \frac{1}{zL}\sum_{l}\frac{1}{1+\frac{s_\alpha}{z}l^{\alpha-1}}
\end{align}
Then, keeping only the leading order contribution close to the accumulation point $l\gg 1$, we find
\begin{align}
    G(\omega) \approx \frac{1}{z}\left[1-\frac{s_\alpha}{zL}\sum_{l}\frac{1}{l^{1-\alpha}}+\mathcal{O}(L^{-1})\right] \approx \frac{1}{z} +\mathcal{O}(L^{\alpha-1}).
\end{align}
It follows that, due to the peculiar properties of the single particle spectrum in the strong long-range regime, the green function satisfies the same equation as in the zero dimensional SYK model without any hopping contribution, apart from finite size corrections which are subleading in the $L\to\infty$ limit.
\section{Thermodynamics}
 The thermodynamics of the system is derived by first computing its free energy. In the $N,L\to\infty$ limit, limit, this is obtained by evaluating the action in Eq.,\eqref{eq: action} at the saddle point solutions for finite temperatures. The free energy then takes the form 
\begin{align}
    F = -\frac{1}{\beta L}\sum_{n}\sum_{k}\ln\left(-i\omega_n+\varepsilon_k+\Sigma(i\omega_n))\right)-\frac{U^2}{2q}\int_0^\beta d\tau\left[G^q(\beta-\tau)G^q(\tau)+\Sigma(\tau)G(\beta-\tau)\right],\label{eq: free energy}
\end{align}
or equivalently, 
\begin{align}
    F = -\frac{1}{L}\sum_k\int_0^\beta d\tau\ln[(\partial_\tau+\varepsilon_k)\delta(\tau)+\Sigma(\tau)]-\frac{U^2}{2q}\int_0^\beta d\tau\left[G^q(\beta-\tau)G^q(\tau)+\Sigma(\tau)G(\beta-\tau)\right].
\end{align}
Using Eq.,\eqref{eq: saddle point equation} and switching from the integral over $\tau$ in the second term of the above expression to a sum over Matsubara frequencies, the free energy simplifies to 
\begin{align}
    F = -\frac{1}{\beta L}\sum_{n}\sum_{k}\ln\left(-i\omega_n+\varepsilon_k+\Sigma(i\omega_n))\right)-\left(\frac{2q-1}{2q}\right)\frac{1}{\beta}\sum_{n}\Sigma(i\omega_n)G(i\omega_n).\label{eq: free energy matsubara}
\end{align}
\subsection{Weak long-range and short-range regimes}
When $\alpha>1$,  the sum over the Fourier modes $k$ can be replaced by an integral over the energy $\varepsilon$ using the density of states  $g(\varepsilon)$, leading to
\begin{align}
    F = -\frac{1}{\beta}\sum_{n}\int d\varepsilon g(\varepsilon)\ln\left(-i\omega_n+\varepsilon+\Sigma(i\omega_n))\right)-\left(\frac{2q-1}{2q}\right)\frac{1}{\beta}\sum_{n}\Sigma(i\omega_n)G(i\omega_n).\label{eq: free energy weak LR}
\end{align}
In the NFL regime $\alpha>\alpha_c(q) = q+1/2$, following the prescription introduced in Ref. \cite{haldar2018higher}, we assume the reparameterization symmetry to approximately hold. Then we find the finite temperature solutions for $G$ and $\Sigma$ by mapping $\tau = f(\sigma)$ such that the bilocal fields transform as
\begin{align}
    Q(\tau_1,\tau_2)\to \tilde{Q}(\tau_1,\tau_2) =\left[f'(\tau_1)f'(\tau_2)\right]^{\Delta_Q}Q(f(\tau_1),f(\tau_2)),
\end{align}
where $Q = G,\Sigma$ and $\Delta_Q = \Delta,\Delta_\Sigma$. Specifically, the infinite line $-\infty<\tau<+\infty$ at $T=0$ is mapped to $0<\tau<\beta$ by the reparametrization
$\tau\to (\beta/\pi)\tan(\pi\tau/\beta)$.  Applying this transformation to the zero-temperature solutions yields the finite temperature scaling forms for $G$ and $\Sigma$, and of their spectral densities. 

Inserting these into Eq.\,\eqref{eq: free energy matsubara}, along with the low-energy behavior of the density of states  $g(\varepsilon)\approx g_\alpha|\varepsilon
|^{-\gamma_\alpha}$ , and keeping only the leading order term in the low-temperature limit $T\to 0$ ($\beta\to\infty$), we find 
\begin{align}
 F(T)\approx T^{(2\Delta_\Sigma-1)(1-\gamma_\alpha)+1} = T^{(2(2q-1)\Delta-1)(1-\gamma_\alpha)+1}.   
\end{align}
Using the $\alpha$-dependent values of $\Delta$ and $\gamma_\alpha$ from Eqs.\,\eqref{eq: Delta NFL} and\,\eqref{eq: gamma_alpha g_alpha}, we obtain $F(T)\approx T^{\zeta+1}$ where 
\begin{align}
    \zeta =  \frac{(2(2q-1)\Delta-1)}{\alpha-1} = \begin{cases}
    \frac{4-q}{1+2q}  &\alpha>3\\\\
    \frac{4q-10-2\alpha(2q-3)}{\left[1+2q(\alpha-2)\right](\alpha-1)} &\alpha_c(q)<\alpha<3
    \end{cases}.
\end{align}
Thus, the low-temperature entropy scaling is  $S(T) = -\partial F/\partial T\approx T^{\zeta}$.

In the FL regime $1<\alpha<\alpha_c(q)$, the self energy contribution in Eq.\,\eqref{eq: free energy weak LR} can be neglected, leading to
\begin{align}
    F\approx -\frac{1}{\beta}\sum_{n}\int d\varepsilon g(\varepsilon)\ln\left(-i\omega_n+\varepsilon\right)
\end{align}
Performing the Matsubara sum and using the low-energy expansion of the density of states, we find
\begin{align}
    F\approx -\frac{1}{\beta}\int d\varepsilon g_\alpha|\varepsilon|^{\frac{1}{\alpha-1}-1}\ln\left(1+e^{\beta\varepsilon}\right).
\end{align}
Then, changing variables to $y = \beta\varepsilon$ in the integral and keeping only the leading term in the low-temperature limit $T\to 0$ ($\beta\to\infty$),  we get $F\approx T^{1+\frac{1}{\alpha-1}}$, leading to an entropy exponent $\zeta =(\alpha-1)^{-1}$.
\subsection{Strong long-range regime}
In the strong long-range case $0\leq\alpha<1$ we start from the expression for the free energy in Eq.\,\eqref{eq: free energy matsubara}, however in this case we can not pass from a sum over the Fourier modes to a continuum integral as the spectrum is discrete also in the thermodynamic limit. Instead we insert the expression for the discrete spectrum $\lim_{L\to\infty}\varepsilon_k = \varepsilon_l$ in the first term of Eq.\,\eqref{eq: free energy matsubara} and we keep the sum over the integer label $l$ leading to 
\begin{align}
\frac{1}{\beta L}\sum_n\sum_k\ln(-i\omega_n+\varepsilon_k+\Sigma(i\omega_n))\approx\frac{1}{\beta}\sum_{n}\ln(-i\omega_n+\Sigma(i\omega_n)) +\frac{1}{\beta L}\sum_{n}\sum_{l}\ln(1+\frac{\varepsilon_l}{\Sigma(i\omega_n)-i\omega_n}).\label{eq: free energy discrete}
\end{align}
Since the spectrum accumulates at $l\to\infty$, it follows that the leading contribution to the sum over $l$ will come from that the values of $\varepsilon_l$ close to the accumulation point. Accordingly we substitute the $\varepsilon_l$ in Eq.\eqref{eq: free energy discrete} with it expansion at large $l$, i.e., $\varepsilon_l\approx s_\alpha l^{\alpha-1}$, leading to
\begin{align}
\frac{1}{\beta L}\sum_n\sum_k\ln(-i\omega_n+\varepsilon_k+\Sigma(i\omega_n))\approx\frac{1}{\beta}\sum_{n}\ln(-i\omega_n+\Sigma(i\omega_n)) +\frac{1}{\beta L}\sum_{n}\sum_{l}\ln(1+\frac{s_\alpha}{(\Sigma(i\omega_n)-i\omega_n)l^{1-\alpha}}).
\end{align}
Then keeping only the leading order contribution as $l\to\infty$ we find
\begin{align}
\frac{1}{\beta L}\sum_n\sum_k\ln(-i\omega_n+\varepsilon_k+\Sigma(i\omega_n))\approx \frac{1}{\beta}\sum_{n}\ln(-i\omega_n+\Sigma(i\omega_n)) +\frac{1}{\beta}\sum_{n}\frac{1}{(\Sigma(i\omega_n)-i\omega_n)}\frac{1}{L}\sum_{l}\frac{s_\alpha}{l^{1-\alpha}}
\end{align}
Noticing that in the thermodynamic limit $L\to\infty$ we have that $\frac{1}{L}\sum_{l}\frac{s_\alpha}{l^{1-\alpha}} = \mathcal{O}(L^{\alpha-1})$, we obtain the following result for the system free energy
\begin{align}
    F = F_{\mathrm{SYK}} + \mathcal{O}(L^{\alpha-1}),
\end{align}
where 
\begin{align}
    F_{\mathrm{SYK}} = \frac{1}{\beta}\sum_{n}\ln(-i\omega_n+\Sigma(i\omega_n))-\left(\frac{2q-1}{2q}\right)\frac{1}{\beta}\sum_{n}\Sigma(i\omega_n)G(i\omega_n),
\end{align}
is the free energy of the standard zero-dimensional SYK model. Finally the entropy is obtained by differentiating with respect to the temperature leading to
\begin{align}
    S = -\frac{\partial F}{\partial T} = S_{\mathrm{SYK}} + \mathcal{O}(L^{\alpha-1}).
\end{align}
This tells us that, in the strong long-range regime, the system dysplays the same free energy and therefore the same entropy as the standard SYK model up to subleading finite size corrections which vanish in the thermodynamic limit as $\mathcal{O}(L^{\alpha-1})$. 
\section{Additional details on the numerical results}
\subsection{Density of states}
\begin{figure*}
    \centering    \includegraphics[width=\linewidth]{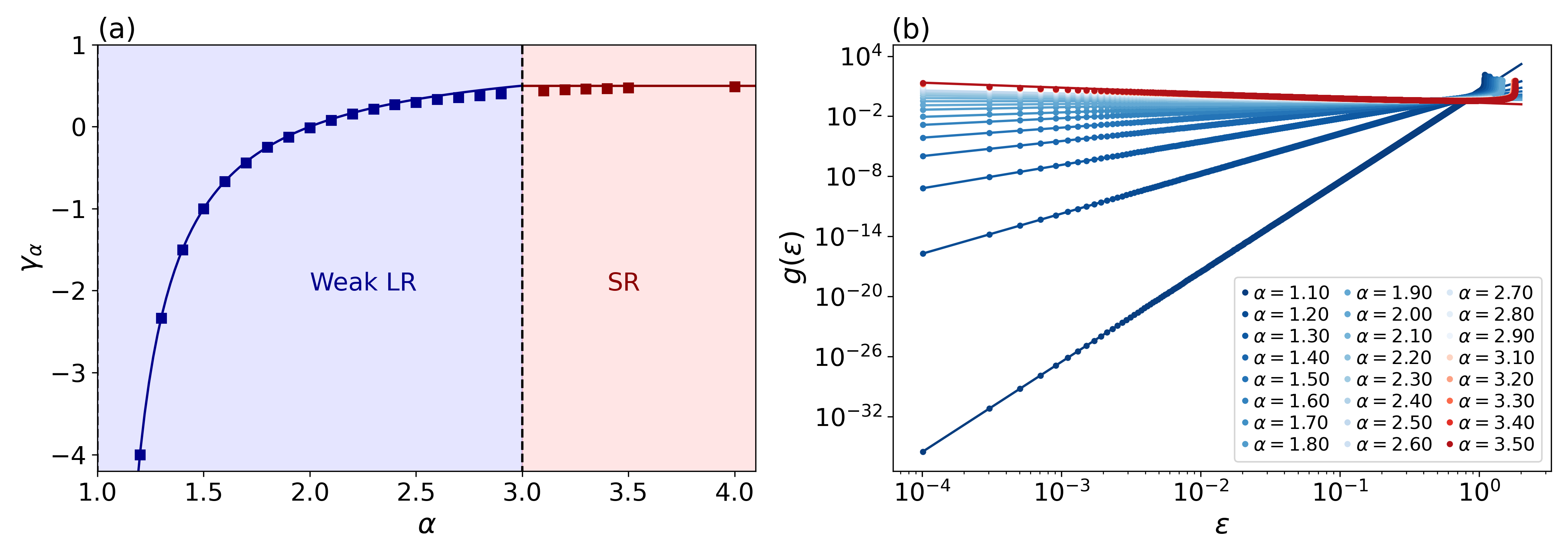}
    \caption{(a) Comparison between the numerically obtained dispersion relation exponent $\gamma_\alpha$ (squares) and the analytical prediction from Eq.\,\eqref{eq: dispersion relation} (solid line), plotted as a function of the interaction range parameter. The vertical dashed line marks the transition between the weak LR and SR regimes at $\alpha = 3$. (b) Numerically computed density of states $g(\varepsilon)$ (scatter plots) as a function of energy $\varepsilon$, compared with the low-energy power-law fits (solid curves) for different values of $\alpha$. The blue curves correspond to the weak long-range (LR) regime, while the red curves correspond to the short-range (SR) regime.}
   \label{fig: gamma_g}
\end{figure*}
To numerically solve the saddle point equations for the system’s Green function and self-energy, the first step is to compute the density of states, $g(\varepsilon)$, corresponding to the long-range single-particle spectrum $\varepsilon_\alpha(k) = \mu - f_\alpha(k)$. This is defined as
\begin{align}
    g(\varepsilon) = \int_{-\pi}^\pi \frac{dk}{2\pi}\delta(\varepsilon-\varepsilon_\alpha (k)).
\end{align}

For accurate analysis, particularly to capture the correct fermionic scaling dimension, it is essential that the density of states exhibits the appropriate low-energy behavior. To this end, we numerically computed $g(\varepsilon)$  based on the long-range single-particle spectrum, carefully examining its low-energy characteristics. We then imposed the correct asymptotic behavior, which is expressed as
\begin{align}         g(\varepsilon)\approx\begin{cases}\varepsilon^{\frac{1}{\alpha-1}-1} &1<\alpha<3\\
    \varepsilon^{-1/2}&\alpha> 3
    \end{cases}\label{eq: dispersion relation}
\end{align} 

Figure \ref{fig: gamma_g}(a) compares the numerically obtained dispersion relation exponent $\gamma_\alpha$ determined by fitting the computed density of states to the power-law form $g(\varepsilon) =  g_\alpha\varepsilon^{-\gamma_\alpha}$ , with the analytical predictions from Eq.\,\eqref{eq: dispersion relation}, plotted as a function of $\alpha$. This comparison serves to benchmark the agreement between the numerical results and theoretical expectations.

In Figure \ref{fig: gamma_g}(b), we show the numerically computed density of states (scatter plots) as a function of energy $\varepsilon$, along with the corresponding low-energy power-law fits for different values of $\alpha$. The blue curves represent the results in the weak LR ($1<\alpha<3$) regime, while the red curves correspond to the SR regime ($\alpha>3$). 
\subsection{Solving the saddle point equations}
To numerically solve the saddle-point equations [Eqs. (6) and (7), main text], we begin by initializing the Green's function $G(i\omega_n)$ with a reasonable guess. In our case, we use the non-interacting Green's function $G_0(i\omega_n) = 1/(i\omega_n + \mu)$. as the starting point. We then compute its Fourier transform $G(\tau)$ in the immaginary time domain through the Matsubara summation
\begin{align}
    G(\tau) = \frac{1}{\beta}\sum_n \left[G(i\omega_n)-\frac{1}{i\omega_n}\right]e^{-i\omega_n\tau}-\frac{1}{2},
\end{align}
where the non-interacting contribution has been subtracted and added back. This procedure reduces Gibb's oscillations near the endpoints of the Fourier transform, making the numerical computation more stable and controlled. This is then used to compute the self-energy $\Sigma(\tau)$ as
\begin{align}
    \Sigma(\tau) &= (-1)^{q+1}U^2G^q(\tau)G^{q-1}(-\tau).
\end{align}
Finally, we transform $\Sigma(\tau)$ to $\Sigma(i\omega_n)$ via the inverse Fourier transform
\begin{align}
    \Sigma(i\omega_n) = \int_0^\beta d\tau\Sigma(\tau)e^{i\omega_n\tau}
\end{align}
and we compute the new $G(i\omega_n)$ as
\begin{align}
    G_\mathrm{new}(i\omega_n) &= aG_\mathrm{old}(i\omega_n)+(1-a)\int_{-\Lambda}^{\Lambda}d\varepsilon\frac{g(\varepsilon)}{i\omega_n-\varepsilon-\Sigma(i\omega_n)},\label{eq: green function},
\end{align}
where we fed back a part of the old G along with the usual expression in order to speed up the iteration convergence. This closes the iterative loop.  We iterate this process till the difference between the new $G(i\omega_n)$ and the
old one has become desirably small.

Figure \ref{fig: Green_Functions} shows the immaginary part of the Green function $\rho(\omega) = -\Im G(\omega)/\pi$ as a function of $\omega$, for the smallest temperature we considered in the FL regime $\alpha<q+1/2 = 2/5$ (panel (a)) and in the NFL regime $\alpha>q+1/2 = 2/5$ (panel (b)). The numerical results (scatter plots) are compared with the power law fit $\rho(\omega)\propto \omega^{2\Delta-1}$ at low $\omega$ we used to extract the fermionic scaling dimension $\Delta$ for different values of $\alpha$ (dashed lines). The results show excellent agreement with analytical expectations.
\begin{figure*}
    \centering    \includegraphics[width=\linewidth]{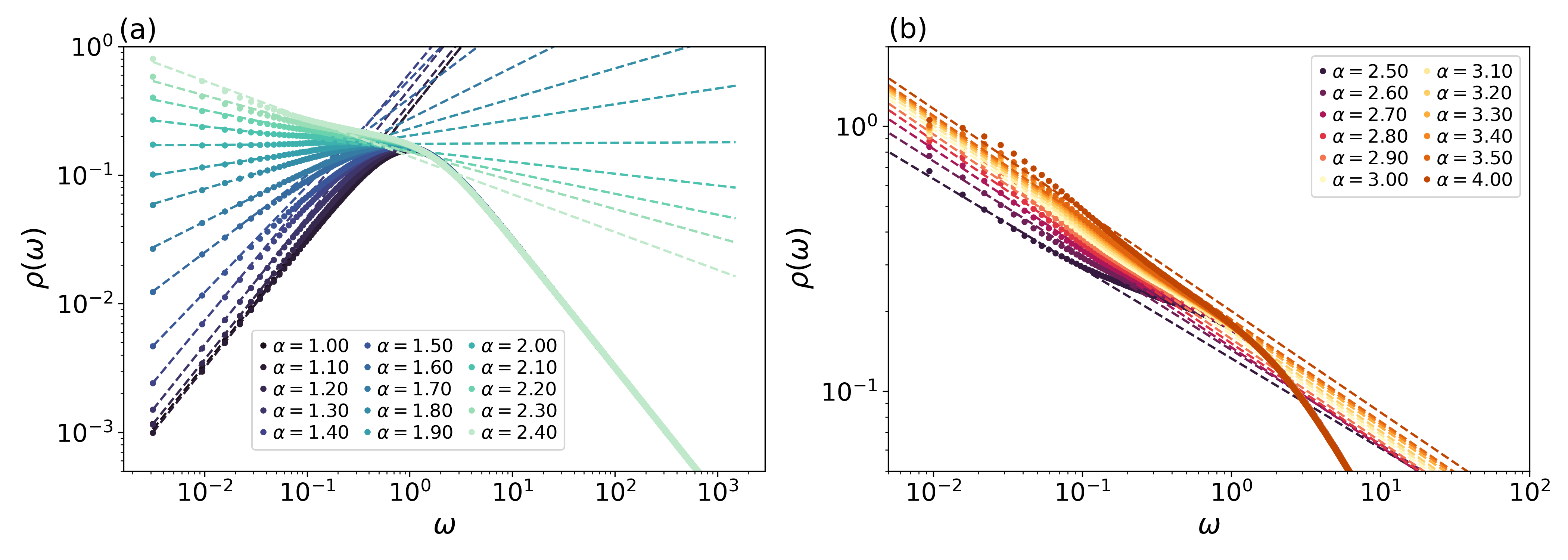}
    \caption{Immaginary part of the Green function $\rho(\omega) = -\mathrm{Im}G(\omega)/\pi$ as a function of the Matsubara frequency $\omega$ for different values of $\alpha$ in the NFL regime $1<\alpha<\alpha_c(q = 2) = 5/2$ (panel (a)) and in the FL regime $\alpha>\alpha_c(q = 2) = 5/2$ (panel (b)). The numerical data (dots) are compared with the low-energy power law fits (dashed lines).}
  \label{fig: Green_Functions}
\end{figure*}
\subsection{Free energy and entropy}
In order to ensure the numerical convergence of Eq.\eqref{eq: free energy} it is convenient to introduce a small chemical potential $\mu\approx 0$ and to add and subtract the free Fermi gas contribution reading 
\begin{align}
    F_{\mathrm{free}} = -\frac{1}{\beta}\sum_{n}\ln(-(i\omega_n+\mu)) = -\frac{1}{\beta}\ln(1+e^{\beta\mu}).
\end{align}
Then we obtain
\begin{align}
    F = -\frac{1}{\beta L}\sum_{n}\sum_{k}\ln\left(\frac{i\omega_n+\mu-\varepsilon_k-\Sigma(i\omega_n))}{i\omega_n+\mu}\right)-\frac{U^2}{2q}\int_0^\beta d\tau\left[G^q(\beta-\tau)G^q(\tau)+\Sigma(\tau)G(\beta-\tau)\right]-\frac{1}{\beta}\ln(1+e^{\beta\mu}).
\end{align}
Which can be rewritten using the density of states $g(\varepsilon)$ as
\begin{align}
    F = -\frac{1}{\beta}\sum_{n}\int d\varepsilon g(\varepsilon)\ln\left(\frac{i\omega_n+\mu-\varepsilon-\Sigma(i\omega_n))}{i\omega_n+\mu}\right)-\frac{U^2}{2q}\int_0^\beta d\tau\left[G^q(\beta-\tau)G^q(\tau)+\Sigma(\tau)G(\beta-\tau)\right]-\frac{1}{\beta}\ln(1+e^{\beta\mu}).
\end{align}
Finally, the entropy $S=-\partial F/\partial T$ is evaluated by computing numerical derivative of $F(T)$.

\end{widetext}

%%%%%%%%%%%%%%%%%%%%%%%%%%%%%%%%%%%%%%%

%%%%%%%%%%
%\bibliography{Bibliography.bib}

\begin{thebibliography}{41}%
\makeatletter
\providecommand \@ifxundefined [1]{%
 \@ifx{#1\undefined}
}%
\providecommand \@ifnum [1]{%
 \ifnum #1\expandafter \@firstoftwo
 \else \expandafter \@secondoftwo
 \fi
}%
\providecommand \@ifx [1]{%
 \ifx #1\expandafter \@firstoftwo
 \else \expandafter \@secondoftwo
 \fi
}%
\providecommand \natexlab [1]{#1}%
\providecommand \enquote  [1]{``#1''}%
\providecommand \bibnamefont  [1]{#1}%
\providecommand \bibfnamefont [1]{#1}%
\providecommand \citenamefont [1]{#1}%
\providecommand \href@noop [0]{\@secondoftwo}%
\providecommand \href [0]{\begingroup \@sanitize@url \@href}%
\providecommand \@href[1]{\@@startlink{#1}\@@href}%
\providecommand \@@href[1]{\endgroup#1\@@endlink}%
\providecommand \@sanitize@url [0]{\catcode `\\12\catcode `\$12\catcode `\&12\catcode `\#12\catcode `\^12\catcode `\_12\catcode `\%12\relax}%
\providecommand \@@startlink[1]{}%
\providecommand \@@endlink[0]{}%
\providecommand \url  [0]{\begingroup\@sanitize@url \@url }%
\providecommand \@url [1]{\endgroup\@href {#1}{\urlprefix }}%
\providecommand \urlprefix  [0]{URL }%
\providecommand \Eprint [0]{\href }%
\providecommand \doibase [0]{https://doi.org/}%
\providecommand \selectlanguage [0]{\@gobble}%
\providecommand \bibinfo  [0]{\@secondoftwo}%
\providecommand \bibfield  [0]{\@secondoftwo}%
\providecommand \translation [1]{[#1]}%
\providecommand \BibitemOpen [0]{}%
\providecommand \bibitemStop [0]{}%
\providecommand \bibitemNoStop [0]{.\EOS\space}%
\providecommand \EOS [0]{\spacefactor3000\relax}%
\providecommand \BibitemShut  [1]{\csname bibitem#1\endcsname}%
\let\auto@bib@innerbib\@empty
%</preamble>
\bibitem [{\citenamefont {Hawking}(1976)}]{hawking1976breakdown}%
  \BibitemOpen
  \bibfield  {author} {\bibinfo {author} {\bibfnamefont {S.~W.}\ \bibnamefont {Hawking}},\ }\bibfield  {title} {\bibinfo {title} {Breakdown of predictability in gravitational collapse},\ }\href {https://doi.org/10.1103/PhysRevD.14.2460} {\bibfield  {journal} {\bibinfo  {journal} {Phys. Rev. D}\ }\textbf {\bibinfo {volume} {14}},\ \bibinfo {pages} {2460} (\bibinfo {year} {1976})}\BibitemShut {NoStop}%
\bibitem [{\citenamefont {Susskind}(1995)}]{susskind1995theworld}%
  \BibitemOpen
  \bibfield  {author} {\bibinfo {author} {\bibfnamefont {L.}~\bibnamefont {Susskind}},\ }\bibfield  {title} {\bibinfo {title} {{The world as a hologram}},\ }\href {https://doi.org/10.1063/1.531249} {\bibfield  {journal} {\bibinfo  {journal} {Journal of Mathematical Physics}\ }\textbf {\bibinfo {volume} {36}},\ \bibinfo {pages} {6377} (\bibinfo {year} {1995})}\BibitemShut {NoStop}%
\bibitem [{\citenamefont {Franz}\ and\ \citenamefont {Rozali}(2018)}]{franz2018mimicking}%
  \BibitemOpen
  \bibfield  {author} {\bibinfo {author} {\bibfnamefont {M.}~\bibnamefont {Franz}}\ and\ \bibinfo {author} {\bibfnamefont {M.}~\bibnamefont {Rozali}},\ }\bibfield  {title} {\bibinfo {title} {Mimicking black hole event horizons in atomic and solid-state systems},\ }\href {https://doi.org/10.1038/s41578-018-0058-z} {\bibfield  {journal} {\bibinfo  {journal} {Nature Reviews Materials}\ }\textbf {\bibinfo {volume} {3}},\ \bibinfo {pages} {491} (\bibinfo {year} {2018})}\BibitemShut {NoStop}%
\bibitem [{\citenamefont {Hartnoll}\ \emph {et~al.}(2018)\citenamefont {Hartnoll}, \citenamefont {Lucas},\ and\ \citenamefont {Sachdev}}]{hartnoll2018holographic}%
  \BibitemOpen
  \bibfield  {author} {\bibinfo {author} {\bibfnamefont {S.}~\bibnamefont {Hartnoll}}, \bibinfo {author} {\bibfnamefont {A.}~\bibnamefont {Lucas}},\ and\ \bibinfo {author} {\bibfnamefont {S.}~\bibnamefont {Sachdev}},\ }\href {https://books.google.de/books?id=DW9RDwAAQBAJ} {\emph {\bibinfo {title} {Holographic Quantum Matter}}},\ The MIT Press\ (\bibinfo  {publisher} {MIT Press},\ \bibinfo {year} {2018})\BibitemShut {NoStop}%
\bibitem [{\citenamefont {Strominger}\ and\ \citenamefont {Vafa}(1996)}]{strominger1996microscopic}%
  \BibitemOpen
  \bibfield  {author} {\bibinfo {author} {\bibfnamefont {A.}~\bibnamefont {Strominger}}\ and\ \bibinfo {author} {\bibfnamefont {C.}~\bibnamefont {Vafa}},\ }\bibfield  {title} {\bibinfo {title} {Microscopic origin of the bekenstein-hawking entropy},\ }\href {https://doi.org/https://doi.org/10.1016/0370-2693(96)00345-0} {\bibfield  {journal} {\bibinfo  {journal} {Physics Letters B}\ }\textbf {\bibinfo {volume} {379}},\ \bibinfo {pages} {99} (\bibinfo {year} {1996})}\BibitemShut {NoStop}%
\bibitem [{\citenamefont {Shenker}\ and\ \citenamefont {Stanford}(2014)}]{shenker2014blackholes}%
  \BibitemOpen
  \bibfield  {author} {\bibinfo {author} {\bibfnamefont {S.~H.}\ \bibnamefont {Shenker}}\ and\ \bibinfo {author} {\bibfnamefont {D.}~\bibnamefont {Stanford}},\ }\bibfield  {title} {\bibinfo {title} {Black holes and the butterfly effect},\ }\href {https://doi.org/10.1007/JHEP03(2014)067} {\bibfield  {journal} {\bibinfo  {journal} {Journal of High Energy Physics}\ }\textbf {\bibinfo {volume} {2014}},\ \bibinfo {pages} {67} (\bibinfo {year} {2014})}\BibitemShut {NoStop}%
\bibitem [{\citenamefont {Swingle}\ \emph {et~al.}(2016)\citenamefont {Swingle}, \citenamefont {Bentsen}, \citenamefont {Schleier-Smith},\ and\ \citenamefont {Hayden}}]{swingle2016measuring}%
  \BibitemOpen
  \bibfield  {author} {\bibinfo {author} {\bibfnamefont {B.}~\bibnamefont {Swingle}}, \bibinfo {author} {\bibfnamefont {G.}~\bibnamefont {Bentsen}}, \bibinfo {author} {\bibfnamefont {M.}~\bibnamefont {Schleier-Smith}},\ and\ \bibinfo {author} {\bibfnamefont {P.}~\bibnamefont {Hayden}},\ }\bibfield  {title} {\bibinfo {title} {Measuring the scrambling of quantum information},\ }\href {https://doi.org/10.1103/PhysRevA.94.040302} {\bibfield  {journal} {\bibinfo  {journal} {Phys. Rev. A}\ }\textbf {\bibinfo {volume} {94}},\ \bibinfo {pages} {040302} (\bibinfo {year} {2016})}\BibitemShut {NoStop}%
\bibitem [{\citenamefont {Hawking}(1975)}]{hawking1975particle}%
  \BibitemOpen
  \bibfield  {author} {\bibinfo {author} {\bibfnamefont {S.~W.}\ \bibnamefont {Hawking}},\ }\bibfield  {title} {\bibinfo {title} {Particle creation by black holes},\ }\href {https://doi.org/10.1007/BF02345020} {\bibfield  {journal} {\bibinfo  {journal} {Communications in Mathematical Physics}\ }\textbf {\bibinfo {volume} {43}},\ \bibinfo {pages} {199} (\bibinfo {year} {1975})}\BibitemShut {NoStop}%
\bibitem [{\citenamefont {Bekenstein}(1973)}]{bekenstein1973blackholes}%
  \BibitemOpen
  \bibfield  {author} {\bibinfo {author} {\bibfnamefont {J.~D.}\ \bibnamefont {Bekenstein}},\ }\bibfield  {title} {\bibinfo {title} {Black holes and entropy},\ }\href {https://doi.org/10.1103/PhysRevD.7.2333} {\bibfield  {journal} {\bibinfo  {journal} {Phys. Rev. D}\ }\textbf {\bibinfo {volume} {7}},\ \bibinfo {pages} {2333} (\bibinfo {year} {1973})}\BibitemShut {NoStop}%
\bibitem [{\citenamefont {Bardeen}\ \emph {et~al.}(1973)\citenamefont {Bardeen}, \citenamefont {Carter},\ and\ \citenamefont {Hawking}}]{bardeen1973thefour}%
  \BibitemOpen
  \bibfield  {author} {\bibinfo {author} {\bibfnamefont {J.~M.}\ \bibnamefont {Bardeen}}, \bibinfo {author} {\bibfnamefont {B.}~\bibnamefont {Carter}},\ and\ \bibinfo {author} {\bibfnamefont {S.~W.}\ \bibnamefont {Hawking}},\ }\bibfield  {title} {\bibinfo {title} {The four laws of black hole mechanics},\ }\href {https://doi.org/10.1007/BF01645742} {\bibfield  {journal} {\bibinfo  {journal} {Communications in Mathematical Physics}\ }\textbf {\bibinfo {volume} {31}},\ \bibinfo {pages} {161} (\bibinfo {year} {1973})}\BibitemShut {NoStop}%
\bibitem [{\citenamefont {Sachdev}\ and\ \citenamefont {Ye}(1993)}]{sachdev1993gapless}%
  \BibitemOpen
  \bibfield  {author} {\bibinfo {author} {\bibfnamefont {S.}~\bibnamefont {Sachdev}}\ and\ \bibinfo {author} {\bibfnamefont {J.}~\bibnamefont {Ye}},\ }\bibfield  {title} {\bibinfo {title} {Gapless spin-fluid ground state in a random quantum heisenberg magnet},\ }\href {https://doi.org/10.1103/PhysRevLett.70.3339} {\bibfield  {journal} {\bibinfo  {journal} {Phys. Rev. Lett.}\ }\textbf {\bibinfo {volume} {70}},\ \bibinfo {pages} {3339} (\bibinfo {year} {1993})}\BibitemShut {NoStop}%
\bibitem [{\citenamefont {Kitaev}(2015)}]{kitaev2015simple}%
  \BibitemOpen
  \bibfield  {author} {\bibinfo {author} {\bibfnamefont {A.}~\bibnamefont {Kitaev}},\ }\bibfield  {title} {\bibinfo {title} {A simple model of quantum holography},\ }\href {https://online.kitp.ucsb.edu/online/entangled15/kitaev/} {\bibfield  {journal} {\bibinfo  {journal} {in KITP Strings Seminar and Entanglement 2015 Program}\ } (\bibinfo {year} {2015})}\BibitemShut {NoStop}%
\bibitem [{\citenamefont {Maldacena}\ and\ \citenamefont {Stanford}(2016)}]{maldacena2016remarks}%
  \BibitemOpen
  \bibfield  {author} {\bibinfo {author} {\bibfnamefont {J.}~\bibnamefont {Maldacena}}\ and\ \bibinfo {author} {\bibfnamefont {D.}~\bibnamefont {Stanford}},\ }\bibfield  {title} {\bibinfo {title} {Remarks on the sachdev-ye-kitaev model},\ }\href {https://doi.org/10.1103/PhysRevD.94.106002} {\bibfield  {journal} {\bibinfo  {journal} {Phys. Rev. D}\ }\textbf {\bibinfo {volume} {94}},\ \bibinfo {pages} {106002} (\bibinfo {year} {2016})}\BibitemShut {NoStop}%
\bibitem [{\citenamefont {Parcollet}\ and\ \citenamefont {Georges}(1999)}]{parcollet1999nonfermi}%
  \BibitemOpen
  \bibfield  {author} {\bibinfo {author} {\bibfnamefont {O.}~\bibnamefont {Parcollet}}\ and\ \bibinfo {author} {\bibfnamefont {A.}~\bibnamefont {Georges}},\ }\bibfield  {title} {\bibinfo {title} {Non-fermi-liquid regime of a doped mott insulator},\ }\href {https://doi.org/10.1103/PhysRevB.59.5341} {\bibfield  {journal} {\bibinfo  {journal} {Phys. Rev. B}\ }\textbf {\bibinfo {volume} {59}},\ \bibinfo {pages} {5341} (\bibinfo {year} {1999})}\BibitemShut {NoStop}%
\bibitem [{\citenamefont {Chowdhury}\ \emph {et~al.}(2022)\citenamefont {Chowdhury}, \citenamefont {Georges}, \citenamefont {Parcollet},\ and\ \citenamefont {Sachdev}}]{chowdhury2022syk}%
  \BibitemOpen
  \bibfield  {author} {\bibinfo {author} {\bibfnamefont {D.}~\bibnamefont {Chowdhury}}, \bibinfo {author} {\bibfnamefont {A.}~\bibnamefont {Georges}}, \bibinfo {author} {\bibfnamefont {O.}~\bibnamefont {Parcollet}},\ and\ \bibinfo {author} {\bibfnamefont {S.}~\bibnamefont {Sachdev}},\ }\bibfield  {title} {\bibinfo {title} {Sachdev-ye-kitaev models and beyond: Window into non-fermi liquids},\ }\href {https://doi.org/10.1103/RevModPhys.94.035004} {\bibfield  {journal} {\bibinfo  {journal} {Rev. Mod. Phys.}\ }\textbf {\bibinfo {volume} {94}},\ \bibinfo {pages} {035004} (\bibinfo {year} {2022})}\BibitemShut {NoStop}%
\bibitem [{\citenamefont {Chowdhury}\ \emph {et~al.}(2018)\citenamefont {Chowdhury}, \citenamefont {Werman}, \citenamefont {Berg},\ and\ \citenamefont {Senthil}}]{chowdhury2018translationally}%
  \BibitemOpen
  \bibfield  {author} {\bibinfo {author} {\bibfnamefont {D.}~\bibnamefont {Chowdhury}}, \bibinfo {author} {\bibfnamefont {Y.}~\bibnamefont {Werman}}, \bibinfo {author} {\bibfnamefont {E.}~\bibnamefont {Berg}},\ and\ \bibinfo {author} {\bibfnamefont {T.}~\bibnamefont {Senthil}},\ }\bibfield  {title} {\bibinfo {title} {Translationally invariant non-fermi-liquid metals with critical fermi surfaces: Solvable models},\ }\href {https://doi.org/10.1103/PhysRevX.8.031024} {\bibfield  {journal} {\bibinfo  {journal} {Phys. Rev. X}\ }\textbf {\bibinfo {volume} {8}},\ \bibinfo {pages} {031024} (\bibinfo {year} {2018})}\BibitemShut {NoStop}%
\bibitem [{\citenamefont {Banerjee}\ and\ \citenamefont {Altman}(2017)}]{banerjee2017solvable}%
  \BibitemOpen
  \bibfield  {author} {\bibinfo {author} {\bibfnamefont {S.}~\bibnamefont {Banerjee}}\ and\ \bibinfo {author} {\bibfnamefont {E.}~\bibnamefont {Altman}},\ }\bibfield  {title} {\bibinfo {title} {Solvable model for a dynamical quantum phase transition from fast to slow scrambling},\ }\href {https://doi.org/10.1103/PhysRevB.95.134302} {\bibfield  {journal} {\bibinfo  {journal} {Phys. Rev. B}\ }\textbf {\bibinfo {volume} {95}},\ \bibinfo {pages} {134302} (\bibinfo {year} {2017})}\BibitemShut {NoStop}%
\bibitem [{\citenamefont {Haldar}\ \emph {et~al.}(2018)\citenamefont {Haldar}, \citenamefont {Banerjee},\ and\ \citenamefont {Shenoy}}]{haldar2018higher}%
  \BibitemOpen
  \bibfield  {author} {\bibinfo {author} {\bibfnamefont {A.}~\bibnamefont {Haldar}}, \bibinfo {author} {\bibfnamefont {S.}~\bibnamefont {Banerjee}},\ and\ \bibinfo {author} {\bibfnamefont {V.~B.}\ \bibnamefont {Shenoy}},\ }\bibfield  {title} {\bibinfo {title} {Higher-dimensional sachdev-ye-kitaev non-fermi liquids at lifshitz transitions},\ }\href {https://doi.org/10.1103/PhysRevB.97.241106} {\bibfield  {journal} {\bibinfo  {journal} {Phys. Rev. B}\ }\textbf {\bibinfo {volume} {97}},\ \bibinfo {pages} {241106} (\bibinfo {year} {2018})}\BibitemShut {NoStop}%
\bibitem [{\citenamefont {Jose}\ \emph {et~al.}(2022)\citenamefont {Jose}, \citenamefont {Seo},\ and\ \citenamefont {Uchoa}}]{jose2022nonfermi}%
  \BibitemOpen
  \bibfield  {author} {\bibinfo {author} {\bibfnamefont {G.}~\bibnamefont {Jose}}, \bibinfo {author} {\bibfnamefont {K.}~\bibnamefont {Seo}},\ and\ \bibinfo {author} {\bibfnamefont {B.}~\bibnamefont {Uchoa}},\ }\bibfield  {title} {\bibinfo {title} {Non-fermi liquid behavior in the sachdev-ye-kitaev model for a one-dimensional incoherent semimetal},\ }\href {https://doi.org/10.1103/PhysRevResearch.4.013145} {\bibfield  {journal} {\bibinfo  {journal} {Phys. Rev. Res.}\ }\textbf {\bibinfo {volume} {4}},\ \bibinfo {pages} {013145} (\bibinfo {year} {2022})}\BibitemShut {NoStop}%
\bibitem [{\citenamefont {Sachdev}(2015)}]{sachdev2015bekensteinhawking}%
  \BibitemOpen
  \bibfield  {author} {\bibinfo {author} {\bibfnamefont {S.}~\bibnamefont {Sachdev}},\ }\bibfield  {title} {\bibinfo {title} {Bekenstein-hawking entropy and strange metals},\ }\href {https://doi.org/10.1103/PhysRevX.5.041025} {\bibfield  {journal} {\bibinfo  {journal} {Phys. Rev. X}\ }\textbf {\bibinfo {volume} {5}},\ \bibinfo {pages} {041025} (\bibinfo {year} {2015})}\BibitemShut {NoStop}%
\bibitem [{\citenamefont {Cotler}\ \emph {et~al.}(2017)\citenamefont {Cotler}, \citenamefont {Gur-Ari}, \citenamefont {Hanada}, \citenamefont {Polchinski}, \citenamefont {Saad}, \citenamefont {Shenker}, \citenamefont {Stanford}, \citenamefont {Streicher},\ and\ \citenamefont {Tezuka}}]{cotler2017blackholes}%
  \BibitemOpen
  \bibfield  {author} {\bibinfo {author} {\bibfnamefont {J.~S.}\ \bibnamefont {Cotler}}, \bibinfo {author} {\bibfnamefont {G.}~\bibnamefont {Gur-Ari}}, \bibinfo {author} {\bibfnamefont {M.}~\bibnamefont {Hanada}}, \bibinfo {author} {\bibfnamefont {J.}~\bibnamefont {Polchinski}}, \bibinfo {author} {\bibfnamefont {P.}~\bibnamefont {Saad}}, \bibinfo {author} {\bibfnamefont {S.~H.}\ \bibnamefont {Shenker}}, \bibinfo {author} {\bibfnamefont {D.}~\bibnamefont {Stanford}}, \bibinfo {author} {\bibfnamefont {A.}~\bibnamefont {Streicher}},\ and\ \bibinfo {author} {\bibfnamefont {M.}~\bibnamefont {Tezuka}},\ }\bibfield  {title} {\bibinfo {title} {Black holes and random matrices},\ }\href {https://doi.org/10.1007/JHEP05(2017)118} {\bibfield  {journal} {\bibinfo  {journal} {Journal of High Energy Physics}\ }\textbf {\bibinfo {volume} {2017}},\ \bibinfo {pages} {118} (\bibinfo {year} {2017})}\BibitemShut {NoStop}%
\bibitem [{\citenamefont {Susskind}(2021)}]{susskind2021entanglement}%
  \BibitemOpen
  \bibfield  {author} {\bibinfo {author} {\bibfnamefont {L.}~\bibnamefont {Susskind}},\ }\bibfield  {title} {\bibinfo {title} {Entanglement and chaos in de sitter space holography: An syk example},\ }\href {https://doi.org/10.22128/jhap.2021.455.1005} {\bibfield  {journal} {\bibinfo  {journal} {Journal of Holography Applications in Physics}\ }\textbf {\bibinfo {volume} {1}},\ \bibinfo {pages} {1} (\bibinfo {year} {2021})}\BibitemShut {NoStop}%
\bibitem [{\citenamefont {Maldacena}\ \emph {et~al.}(2016)\citenamefont {Maldacena}, \citenamefont {Shenker},\ and\ \citenamefont {Stanford}}]{maldacena2016abound}%
  \BibitemOpen
  \bibfield  {author} {\bibinfo {author} {\bibfnamefont {J.}~\bibnamefont {Maldacena}}, \bibinfo {author} {\bibfnamefont {S.~H.}\ \bibnamefont {Shenker}},\ and\ \bibinfo {author} {\bibfnamefont {D.}~\bibnamefont {Stanford}},\ }\bibfield  {title} {\bibinfo {title} {A bound on chaos},\ }\href {https://doi.org/10.1007/JHEP08(2016)106} {\bibfield  {journal} {\bibinfo  {journal} {Journal of High Energy Physics}\ }\textbf {\bibinfo {volume} {2016}},\ \bibinfo {pages} {106} (\bibinfo {year} {2016})}\BibitemShut {NoStop}%
\bibitem [{\citenamefont {Fu}\ and\ \citenamefont {Sachdev}(2016)}]{fu2016numerical}%
  \BibitemOpen
  \bibfield  {author} {\bibinfo {author} {\bibfnamefont {W.}~\bibnamefont {Fu}}\ and\ \bibinfo {author} {\bibfnamefont {S.}~\bibnamefont {Sachdev}},\ }\bibfield  {title} {\bibinfo {title} {Numerical study of fermion and boson models with infinite-range random interactions},\ }\href {https://doi.org/10.1103/PhysRevB.94.035135} {\bibfield  {journal} {\bibinfo  {journal} {Phys. Rev. B}\ }\textbf {\bibinfo {volume} {94}},\ \bibinfo {pages} {035135} (\bibinfo {year} {2016})}\BibitemShut {NoStop}%
\bibitem [{\citenamefont {Chamblin}\ \emph {et~al.}(1999)\citenamefont {Chamblin}, \citenamefont {Emparan}, \citenamefont {Johnson},\ and\ \citenamefont {Myers}}]{chamblin1999charged}%
  \BibitemOpen
  \bibfield  {author} {\bibinfo {author} {\bibfnamefont {A.}~\bibnamefont {Chamblin}}, \bibinfo {author} {\bibfnamefont {R.}~\bibnamefont {Emparan}}, \bibinfo {author} {\bibfnamefont {C.~V.}\ \bibnamefont {Johnson}},\ and\ \bibinfo {author} {\bibfnamefont {R.~C.}\ \bibnamefont {Myers}},\ }\bibfield  {title} {\bibinfo {title} {Charged ads black holes and catastrophic holography},\ }\href {https://doi.org/10.1103/PhysRevD.60.064018} {\bibfield  {journal} {\bibinfo  {journal} {Phys. Rev. D}\ }\textbf {\bibinfo {volume} {60}},\ \bibinfo {pages} {064018} (\bibinfo {year} {1999})}\BibitemShut {NoStop}%
\bibitem [{\citenamefont {Faulkner}\ \emph {et~al.}(2011)\citenamefont {Faulkner}, \citenamefont {Liu}, \citenamefont {McGreevy},\ and\ \citenamefont {Vegh}}]{faulkner2011emergent}%
  \BibitemOpen
  \bibfield  {author} {\bibinfo {author} {\bibfnamefont {T.}~\bibnamefont {Faulkner}}, \bibinfo {author} {\bibfnamefont {H.}~\bibnamefont {Liu}}, \bibinfo {author} {\bibfnamefont {J.}~\bibnamefont {McGreevy}},\ and\ \bibinfo {author} {\bibfnamefont {D.}~\bibnamefont {Vegh}},\ }\bibfield  {title} {\bibinfo {title} {Emergent quantum criticality, fermi surfaces, and ${\mathrm{ads}}_{2}$},\ }\href {https://doi.org/10.1103/PhysRevD.83.125002} {\bibfield  {journal} {\bibinfo  {journal} {Phys. Rev. D}\ }\textbf {\bibinfo {volume} {83}},\ \bibinfo {pages} {125002} (\bibinfo {year} {2011})}\BibitemShut {NoStop}%
\bibitem [{\citenamefont {Sen}(2005)}]{sen2005blackhole}%
  \BibitemOpen
  \bibfield  {author} {\bibinfo {author} {\bibfnamefont {A.}~\bibnamefont {Sen}},\ }\bibfield  {title} {\bibinfo {title} {Black hole entropy function and the attractor mechanism in higher derivative gravity},\ }\href {https://doi.org/10.1088/1126-6708/2005/09/038} {\bibfield  {journal} {\bibinfo  {journal} {Journal of High Energy Physics}\ }\textbf {\bibinfo {volume} {2005}},\ \bibinfo {pages} {038} (\bibinfo {year} {2005})}\BibitemShut {NoStop}%
\bibitem [{\citenamefont {Sen}(2008)}]{sen2008entropy}%
  \BibitemOpen
  \bibfield  {author} {\bibinfo {author} {\bibfnamefont {A.}~\bibnamefont {Sen}},\ }\bibfield  {title} {\bibinfo {title} {Entropy function and ads2/cft1 correspondence},\ }\href {https://doi.org/10.1088/1126-6708/2008/11/075} {\bibfield  {journal} {\bibinfo  {journal} {Journal of High Energy Physics}\ }\textbf {\bibinfo {volume} {2008}},\ \bibinfo {pages} {075} (\bibinfo {year} {2008})}\BibitemShut {NoStop}%
\bibitem [{\citenamefont {Wald}(1993)}]{wald1993blackhole}%
  \BibitemOpen
  \bibfield  {author} {\bibinfo {author} {\bibfnamefont {R.~M.}\ \bibnamefont {Wald}},\ }\bibfield  {title} {\bibinfo {title} {Black hole entropy is the noether charge},\ }\href {https://doi.org/10.1103/PhysRevD.48.R3427} {\bibfield  {journal} {\bibinfo  {journal} {Phys. Rev. D}\ }\textbf {\bibinfo {volume} {48}},\ \bibinfo {pages} {R3427} (\bibinfo {year} {1993})}\BibitemShut {NoStop}%
\bibitem [{\citenamefont {Jacobson}\ \emph {et~al.}(1994)\citenamefont {Jacobson}, \citenamefont {Kang},\ and\ \citenamefont {Myers}}]{jacobson1994onblackhole}%
  \BibitemOpen
  \bibfield  {author} {\bibinfo {author} {\bibfnamefont {T.}~\bibnamefont {Jacobson}}, \bibinfo {author} {\bibfnamefont {G.}~\bibnamefont {Kang}},\ and\ \bibinfo {author} {\bibfnamefont {R.~C.}\ \bibnamefont {Myers}},\ }\bibfield  {title} {\bibinfo {title} {On black hole entropy},\ }\href {https://doi.org/10.1103/PhysRevD.49.6587} {\bibfield  {journal} {\bibinfo  {journal} {Phys. Rev. D}\ }\textbf {\bibinfo {volume} {49}},\ \bibinfo {pages} {6587} (\bibinfo {year} {1994})}\BibitemShut {NoStop}%
\bibitem [{\citenamefont {Parcollet}\ \emph {et~al.}(1998)\citenamefont {Parcollet}, \citenamefont {Georges}, \citenamefont {Kotliar},\ and\ \citenamefont {Sengupta}}]{parcollet1998overscreened}%
  \BibitemOpen
  \bibfield  {author} {\bibinfo {author} {\bibfnamefont {O.}~\bibnamefont {Parcollet}}, \bibinfo {author} {\bibfnamefont {A.}~\bibnamefont {Georges}}, \bibinfo {author} {\bibfnamefont {G.}~\bibnamefont {Kotliar}},\ and\ \bibinfo {author} {\bibfnamefont {A.}~\bibnamefont {Sengupta}},\ }\bibfield  {title} {\bibinfo {title} {Overscreened multichannel $\mathrm{SU}(n)$ kondo model: Large-$n$ solution and conformal field theory},\ }\href {https://doi.org/10.1103/PhysRevB.58.3794} {\bibfield  {journal} {\bibinfo  {journal} {Phys. Rev. B}\ }\textbf {\bibinfo {volume} {58}},\ \bibinfo {pages} {3794} (\bibinfo {year} {1998})}\BibitemShut {NoStop}%
\bibitem [{\citenamefont {Georges}\ \emph {et~al.}(2001)\citenamefont {Georges}, \citenamefont {Parcollet},\ and\ \citenamefont {Sachdev}}]{parcollet2001quantum}%
  \BibitemOpen
  \bibfield  {author} {\bibinfo {author} {\bibfnamefont {A.}~\bibnamefont {Georges}}, \bibinfo {author} {\bibfnamefont {O.}~\bibnamefont {Parcollet}},\ and\ \bibinfo {author} {\bibfnamefont {S.}~\bibnamefont {Sachdev}},\ }\bibfield  {title} {\bibinfo {title} {Quantum fluctuations of a nearly critical heisenberg spin glass},\ }\href {https://doi.org/10.1103/PhysRevB.63.134406} {\bibfield  {journal} {\bibinfo  {journal} {Phys. Rev. B}\ }\textbf {\bibinfo {volume} {63}},\ \bibinfo {pages} {134406} (\bibinfo {year} {2001})}\BibitemShut {NoStop}%
\bibitem [{\citenamefont {Song}\ \emph {et~al.}(2017)\citenamefont {Song}, \citenamefont {Jian},\ and\ \citenamefont {Balents}}]{song2017strongly}%
  \BibitemOpen
  \bibfield  {author} {\bibinfo {author} {\bibfnamefont {X.-Y.}\ \bibnamefont {Song}}, \bibinfo {author} {\bibfnamefont {C.-M.}\ \bibnamefont {Jian}},\ and\ \bibinfo {author} {\bibfnamefont {L.}~\bibnamefont {Balents}},\ }\bibfield  {title} {\bibinfo {title} {Strongly correlated metal built from sachdev-ye-kitaev models},\ }\href {https://doi.org/10.1103/PhysRevLett.119.216601} {\bibfield  {journal} {\bibinfo  {journal} {Phys. Rev. Lett.}\ }\textbf {\bibinfo {volume} {119}},\ \bibinfo {pages} {216601} (\bibinfo {year} {2017})}\BibitemShut {NoStop}%
\bibitem [{\citenamefont {Defenu}(2021)}]{defenu2021metastability}%
  \BibitemOpen
  \bibfield  {author} {\bibinfo {author} {\bibfnamefont {N.}~\bibnamefont {Defenu}},\ }\bibfield  {title} {\bibinfo {title} {Metastability and discrete spectrum of long-range systems},\ }\href {https://doi.org/10.1073/pnas.2101785118} {\bibfield  {journal} {\bibinfo  {journal} {Proceedings of the National Academy of Sciences}\ }\textbf {\bibinfo {volume} {118}},\ \bibinfo {pages} {e2101785118} (\bibinfo {year} {2021})},\ \Eprint {https://arxiv.org/abs/https://www.pnas.org/doi/pdf/10.1073/pnas.2101785118} {https://www.pnas.org/doi/pdf/10.1073/pnas.2101785118} \BibitemShut {NoStop}%
\bibitem [{\citenamefont {Defenu}\ \emph {et~al.}(2023)\citenamefont {Defenu}, \citenamefont {Donner}, \citenamefont {Macr\`{\i}}, \citenamefont {Pagano}, \citenamefont {Ruffo},\ and\ \citenamefont {Trombettoni}}]{defenu2023longrange}%
  \BibitemOpen
  \bibfield  {author} {\bibinfo {author} {\bibfnamefont {N.}~\bibnamefont {Defenu}}, \bibinfo {author} {\bibfnamefont {T.}~\bibnamefont {Donner}}, \bibinfo {author} {\bibfnamefont {T.}~\bibnamefont {Macr\`{\i}}}, \bibinfo {author} {\bibfnamefont {G.}~\bibnamefont {Pagano}}, \bibinfo {author} {\bibfnamefont {S.}~\bibnamefont {Ruffo}},\ and\ \bibinfo {author} {\bibfnamefont {A.}~\bibnamefont {Trombettoni}},\ }\bibfield  {title} {\bibinfo {title} {Long-range interacting quantum systems},\ }\href {https://doi.org/10.1103/RevModPhys.95.035002} {\bibfield  {journal} {\bibinfo  {journal} {Rev. Mod. Phys.}\ }\textbf {\bibinfo {volume} {95}},\ \bibinfo {pages} {035002} (\bibinfo {year} {2023})}\BibitemShut {NoStop}%
\bibitem [{\citenamefont {Defenu}\ \emph {et~al.}(2024)\citenamefont {Defenu}, \citenamefont {Lerose},\ and\ \citenamefont {Pappalardi}}]{defenu2024outofequilibrium}%
  \BibitemOpen
  \bibfield  {author} {\bibinfo {author} {\bibfnamefont {N.}~\bibnamefont {Defenu}}, \bibinfo {author} {\bibfnamefont {A.}~\bibnamefont {Lerose}},\ and\ \bibinfo {author} {\bibfnamefont {S.}~\bibnamefont {Pappalardi}},\ }\bibfield  {title} {\bibinfo {title} {Out-of-equilibrium dynamics of quantum many-body systems with long-range interactions},\ }\href {https://doi.org/https://doi.org/10.1016/j.physrep.2024.04.005} {\bibfield  {journal} {\bibinfo  {journal} {Physics Reports}\ }\textbf {\bibinfo {volume} {1074}},\ \bibinfo {pages} {1} (\bibinfo {year} {2024})},\ \bibinfo {note} {out-of-equilibrium dynamics of quantum many-body systems with long-range interactions}\BibitemShut {NoStop}%
\bibitem [{\citenamefont {{Kac}}\ \emph {et~al.}(1963)\citenamefont {{Kac}}, \citenamefont {{Uhlenbeck}},\ and\ \citenamefont {{Hemmer}}}]{kac1963van}%
  \BibitemOpen
  \bibfield  {author} {\bibinfo {author} {\bibfnamefont {M.}~\bibnamefont {{Kac}}}, \bibinfo {author} {\bibfnamefont {G.~E.}\ \bibnamefont {{Uhlenbeck}}},\ and\ \bibinfo {author} {\bibfnamefont {P.~C.}\ \bibnamefont {{Hemmer}}},\ }\bibfield  {title} {\bibinfo {title} {{On the van der Waals Theory of the Vapor-Liquid Equilibrium. I. Discussion of a One-Dimensional Model}},\ }\href {https://doi.org/10.1063/1.1703946} {\bibfield  {journal} {\bibinfo  {journal} {Journal of Mathematical Physics}\ }\textbf {\bibinfo {volume} {4}},\ \bibinfo {pages} {216} (\bibinfo {year} {1963})}\BibitemShut {NoStop}%
\bibitem [{\citenamefont {Campa}\ \emph {et~al.}(2014)\citenamefont {Campa}, \citenamefont {Dauxois}, \citenamefont {Fanelli},\ and\ \citenamefont {Ruffo}}]{campa2014physics}%
  \BibitemOpen
  \bibfield  {author} {\bibinfo {author} {\bibfnamefont {A.}~\bibnamefont {Campa}}, \bibinfo {author} {\bibfnamefont {T.}~\bibnamefont {Dauxois}}, \bibinfo {author} {\bibfnamefont {D.}~\bibnamefont {Fanelli}},\ and\ \bibinfo {author} {\bibfnamefont {S.}~\bibnamefont {Ruffo}},\ }\href@noop {} {\emph {\bibinfo {title} {{Physics of Long-Range Interacting Systems}}}}\ (\bibinfo  {publisher} {Oxford Univ. Press},\ \bibinfo {year} {2014})\BibitemShut {NoStop}%
\bibitem [{\citenamefont {Last}(1996)}]{last1996quantum}%
  \BibitemOpen
  \bibfield  {author} {\bibinfo {author} {\bibfnamefont {Y.}~\bibnamefont {Last}},\ }\bibfield  {title} {\bibinfo {title} {Quantum dynamics and decompositions of singular continuous spectra},\ }\href@noop {} {\bibfield  {journal} {\bibinfo  {journal} {Journal of Functional Analysis}\ }\textbf {\bibinfo {volume} {142}},\ \bibinfo {pages} {406} (\bibinfo {year} {1996})}\BibitemShut {NoStop}%
\bibitem [{\citenamefont {Defenu}\ \emph {et~al.}(2019)\citenamefont {Defenu}, \citenamefont {Morigi}, \citenamefont {Dell'Anna},\ and\ \citenamefont {Enss}}]{defenu2019universal}%
  \BibitemOpen
  \bibfield  {author} {\bibinfo {author} {\bibfnamefont {N.}~\bibnamefont {Defenu}}, \bibinfo {author} {\bibfnamefont {G.}~\bibnamefont {Morigi}}, \bibinfo {author} {\bibfnamefont {L.}~\bibnamefont {Dell'Anna}},\ and\ \bibinfo {author} {\bibfnamefont {T.}~\bibnamefont {Enss}},\ }\bibfield  {title} {\bibinfo {title} {Universal dynamical scaling of long-range topological superconductors},\ }\href {https://doi.org/10.1103/PhysRevB.100.184306} {\bibfield  {journal} {\bibinfo  {journal} {Phys. Rev. B}\ }\textbf {\bibinfo {volume} {100}},\ \bibinfo {pages} {184306} (\bibinfo {year} {2019})}\BibitemShut {NoStop}%
\bibitem [{\citenamefont {Abramowitz}\ and\ \citenamefont {Stegun}(1965)}]{abramowitz1964handbook}%
  \BibitemOpen
  \bibfield  {author} {\bibinfo {author} {\bibfnamefont {M.}~\bibnamefont {Abramowitz}}\ and\ \bibinfo {author} {\bibfnamefont {I.}~\bibnamefont {Stegun}},\ }\href {https://books.google.it/books?id=MtU8uP7XMvoC} {\emph {\bibinfo {title} {Handbook of Mathematical Functions: With Formulas, Graphs, and Mathematical Tables}}},\ Applied mathematics series\ (\bibinfo  {publisher} {Dover Publications},\ \bibinfo {year} {1965})\BibitemShut {NoStop}%
\end{thebibliography}
%%%%%%


\end{document}